\documentclass[a4paper,11pt]{article}
\pdfoutput=1

\usepackage{jheppub} 
\usepackage{natbib}
\setcitestyle{square,comma,numbers,sort&compress}
\usepackage{enumerate}
\usepackage{tabularx,booktabs}
\usepackage[dvipsnames]{xcolor}
\usepackage{comment}
\usepackage[caption=false]{subfig}
\usepackage[compat=1.1.0]{tikz-feynman} 
\usepackage{cancel}
\def\mc#1{\mathcal#1}
\usepackage[section]{placeins}
\newcommand{\matc}[1]{  \left(
\begin{array}{c}
 #1
\end{array}
\right)
}

\def\ket#1{\vert\,#1\,\rangle}

\newcolumntype{Y}{>{\centering\arraybackslash}X}
 
\newcommand{\g}[1]{\mathbf{#1}}
\newcommand{\vev}[1]{\langle #1 \rangle}
\newcommand{\gb}[1]{\bar{\mathbf{#1}}}
\newcommand{\red}[1]{\textcolor{black}{#1}}

\definecolor{myRED}{rgb}{0.8, 0.25, 0.33}

\title{Constraining Low-Scale Flavor Models with $\boldsymbol{(g-2)_{\mu}}$ \& Lepton Flavor Violation}

\author[a, b]{M.L. L\'opez-Ib\'a\~nez,}
\author[c]{Aurora Melis,}
\author[d]{M. Jay P\'erez,} 
\author[e]{Moinul Hossain Rahat,} 
\author[f]{and Oscar Vives} 

\affiliation[a]{CAS Key Laboratory of Theoretical Physics, Institute of Theoretical Physics, Chinese Academy of Sciences, Beijing 100190, China}
\affiliation[b]{Departmento de F\'isica, Campus de Rabanales Edif. C2, Universidad de C\'ordoba, E-14071 C\'ordoba, Spain}
\affiliation[c]{Laboratory of High Energy \& Computational Physics, NICPB, R\"avala 10, 10143 Tallinn, Estonia}
\affiliation[d]{Valencia College, Osceola Science Department, Kissimmee, FL 34744, USA}
\affiliation[e]{Institute for Fundamental Theory, Department of Physics,
University of Florida,\\Gainesville, FL 32611, USA }
\affiliation[f]{Departament de Física Tèorica, Universitat de València, Dr. Moliner 50, \\E-46100 Burjassot
\& IFIC, Universitat de València \& CSIC, E-46071, Paterna, Spain}

\emailAdd{maloi2@uv.es}
\emailAdd{aurora.melis@uv.es}
\emailAdd{mperez75@valenciacollege.edu}
\emailAdd{mrahat@ufl.edu}
\emailAdd{oscar.vives@uv.es}

\abstract{{\color{black} We present here two concrete examples of models where a sub-TeV scale breaking of their respective $\mathcal{T}_{13}$ and $A_5$ flavor symmetries is able to account for the recently observed discrepancy in the muon anomalous magnetic moment, $(g-2)_\mu$. Similarities in the flavor structures of the charged-lepton Yukawa matrix and  dipole matrix yielding $(g-2)_\mu$ give rise to strong constraints on low-scale flavor models when bounds from lepton flavor violation (LFV) are imposed. These constraints place stringent limits on the off-diagonal Yukawa structure, suggesting a mostly (quasi-)diagonal texture for models with a low flavor breaking scale $\Lambda_f$. We argue that many of the popular flavor models in the literature designed to explain the fermion masses and mixings are not suitable for reproducing the observed discrepancy in $(g-2)_\mu$, which requires a delicate balance of maintaining a low flavor scale while simultaneously satisfying strong LFV constraints.}}

\begin{document}
\maketitle
\flushbottom

\section{Introduction}

At the dawn of the LHC era, the physics community awaited with bated breath for new electroweak physics whose discovery many were convinced was just around the corner. Defying expectations, more than ten years later and after the ends of Run1 and Run2, a clear sign of new physics has yet to emerge. 

That this is so is especially puzzling, given the common conviction that the Standard Model (SM) cannot be the ultimate theory of “everything” due to the many questions it leaves unanswered. Among others, it can not explain the presence of three families of fermions with different masses and mixings or accommodate the existence of dark matter in the universe. New physics associated with these ``open questions" is certainly required; the only remaining doubt is at what scale this new physics is hiding.

In the case of dark matter, thermally obtaining the correct relic abundance seems to point to the electroweak scale as the scale associated with the dark matter particles, although other mass scales are still possible. In contrast, flavor physics, whose theoretical constraints in the SM come from \emph{dimensionless} Yukawa couplings, does not provide any obvious hints as to the scale of the physics responsible for generating these couplings.

Fortunately, there exist other flavorful couplings, such as dipole moments, with non-vanishing mass dimension that could provide some information on the scale of the new physics responsible for the observed flavor structures in the SM.  Any observation of new contributions to these operators with flavor structures beyond those of the SM Yukawa couplings would most certainly help shed light on the origins of flavor; when coupled with the constraints from other lepton flavor violating (LFV) processes, they may also provide bounds on the masses of new particles responsible for generating them.

Admittedly, deviations in the dipole matrices from their SM predictions may give only indirect evidence for an underlying theory of flavor. New physics satisfying the conservative ansatz of minimal flavor violation (MFV), where the Yukawa couplings are the only source of flavor-changing interactions, can still generate new contributions to the dipole moments beyond those of the SM. 

However, even when the underlying mechanism responsible for generating the flavor structures of the Yukawa and dipole couplings are fundamentally the same, differences in their $\mc O(1)$ coefficients stemming from the details of how they are generated may mean that these matrices are not exactly proportional. 

A framework, using effective operators, for how this could occur was given in Refs. \cite{Calibbi:2020emz, Calibbi:2021qto}. There, the one-loop radiative corrections to the fermion masses, stemming from the breaking of a low-energy flavor symmetry, also generated the dipole matrices. Although the same underlying flavor symmetry controlled the flavor structure of both operators, a differences in their $\mc O(1)$ coefficients arose from the distinct ways of attaching the photon line to the diagrams generating the dipole transitions. 

In such models, the rotation from the flavor to mass basis which diagonalizes the fermion masses would not simultaneously diagonalize the dipole matrices, which may lead to strong constraints from LFV observables. This sensitivity to any deviation between the flavor structures of these two operators make low-energy flavor models that can simultaneously explain both \textcolor{black}{operators successfully} rich targets for study. 

Excitingly, recent experimental results are beginning to hint at just such deviations. \textcolor{black}{The new results from the Muon $g-2$ experiment at Fermilab~\cite{PhysRevLett.126.141801} have confirmed the long-standing discrepancy between the Standard Model (SM) prediction for the anomalous magnetic moment of the muon, $a_\mu = (g_\mu-2)/2$ \cite{Aoyama:2020ynm},\footnote{The community consensus value of muon $g-2$ in the SM is based on latest evaluations of the contributions from quantum electrodynamics (QED) to
tenth order \cite{Aoyama:2012wk, Aoyama:2019ryr}, hadronic vacuum polarization \cite{Davier:2010nc, Davier:2017zfy, Keshavarzi:2018mgv, Colangelo:2018mtw, Hoferichter:2019mqg, Davier:2019can, Keshavarzi:2019abf, Kurz:2014wya}, hadronic light-by-light \cite{Jegerlehner:2017gek, Melnikov:2003xd, Masjuan:2017tvw, Colangelo:2017fiz, Hoferichter:2018kwz, Gerardin:2019vio, Bijnens:2019ghy, Colangelo:2019uex, Pauk:2014rta, Danilkin:2016hnh, Knecht:2018sci, Eichmann:2019bqf, Roig:2019reh, Blum:2019ugy, Colangelo:2014qya}, and electroweak
processes \cite{Jackiw:1972jz, Bars:1972pe, Fujikawa:1972fe, Czarnecki:2002nt, Gnendiger:2013pva}.} and the previous BNL measurements \cite{Brown:2001mga,Muong-2:2006rrc}.}
This outcome makes a strong case for new physics interacting with SM muons at scales not far above the electroweak (EW) scale. On the other hand, a similar 2$\sigma$ tension between theory and experiments for the electron $g-2$ is now under dispute. In the presence of an independent and sufficiently precise measurement of $\alpha$, one can employ $(g-2)_e$ as a test for new physics \cite{Giudice:2012ms}. This has become possible in recent years and the most precise result, obtained by employing matter-wave interferometry with cesium-133 atoms \cite{Parker:2018vye}, highlighted the discrepancy in $(g-2)_e$. Other hints may come from B physics, where similar $3-4\sigma$ deviations are providing hints for violations of lepton universality \cite{Cornella:2021sby}. 

As stated above, a precise determination of the flavor structure of dipole moments would be crucial in determining the mechanism responsible for flavor in the SM. Of course, one possibility is that we find the flavor structure of the dipole matrix proportional to the SM Yukawa couplings, as in a MFV scenario, providing only a lower bound on the scale of flavor-dependent interactions. Nevertheless, given that Yukawa couplings do not restrict the scale of flavor breaking, it is still possible to have a flavor symmetry broken at a low scale, such that the contributions to the dipole moments are sizable. 

 In this work, we follow up on the ideas explored by some of the authors in Refs. \cite{Calibbi:2020emz, Calibbi:2021qto} by providing explicit constructions of flavor models capable of realizing these ideas. As a proof of concept, we build two explicit models with low-energy flavor symmetries based on the groups $\mc T_{13}$ and $A_5$, capable of explaining the muon $g-2$ discrepancy while satisfying all constraints from LFV dipole transitions. We show that the required absence of LFV transitions while maintaining a sizable contribution to the muon anomalous magnetic moment restricts the structure of the flavor symmetry, or, alternatively, the scale of flavor symmetry breaking.

The paper is organized as follows. In Section \ref{sec:review}, we review the salient features of the framework proposed in Refs. \cite{Calibbi:2020emz, Calibbi:2021qto} for constraining low-scale flavor models using the anomalous magnetic moment of the muon and LFV observables in the language of effective field theories (EFT). Sections \ref{sec:T13} and \ref{sec:A5} contain the main results of this work: explicit constructions of low-scale flavor models based on the flavor groups $\mc T_{13}$ and $A_5$ capable of producing sizable contributions to the muons anomalous magnetic moment while evading LFV constraints. We provide some clarifying remarks and comments on the general applicability of such a framework to other flavor models and possible obstacles to extending such an analysis in Section \ref{sec:outlook}. Finally, we summarize and conclude in Section \ref{sec:conclusions}.

\section{Connecting the Charged Lepton Yukawa Structure to $g-2$} \label{sec:review}

We begin by briefly reviewing the idea introduced in Refs. \cite{Calibbi:2020emz, Calibbi:2021qto} that the observed discrepancy in the anomalous magnetic moment of muon can be entirely explained by accounting for contributions from the breaking of a low-energy flavor symmetry. Such a framework works by exploiting the similar flavor structures present in the dipole matrix describing the anomalous magnetic moment, lepton flavor violating transitions and electric dipole moment, and the Yukawa matrix. When both operators are generated from the same underlying theory of flavor, stringent constraints from LFV observables then restrict the off-diagonal entries of the charged lepton Yukawas. 

As is well known, the charged-lepton Yukawa matrix is not completely determined in the Standard Model; the only constraint is that its diagonalization should yield the hierarchical charged lepton mass ratios $y_e/y_\mu = m_e/m_\mu \simeq 0.005$ and $y_\mu/y_\tau = m_\mu/m_\tau \simeq 0.059$. A simple approach for generating such hierarchies is through the spontaneous breaking of an underlying flavor symmetry, transmitted to the SM fermions by heavy messengers \`a la Froggatt-Nielsen (FN) \cite{Froggatt:1978nt,Leurer:1992wg,Leurer:1993gy}. 

Integrating out these heavy messengers yields effective Yukawa interactions of the form 
\begin{equation} \nonumber
\bar{L}\ell H (\varphi_1 \varphi_2 \ldots \varphi_n/\Lambda_f^n ),
\end{equation}
where $\bar{L}$ and $\ell$ are the SM $SU(2)$ doublets and singlets, respectively, the $\varphi_k$ are flavon fields, gauge-singlet scalars transforming non-trivially under the chosen flavor symmetry which encode its breaking, and $\Lambda_f$ is a scale associated with the underlying flavor dynamics. In the electroweak vacuum, such operators yield the mass matrix $v Y_{ij} \bar{L}_i \ell_j$, where $v \simeq 246$ GeV is the Higgs vacuum expectation value (VEV), and the Yukawa matrix $Y_{ij}$ contains the relevant suppression factors $\vev{\varphi_k}/\Lambda_f$. Since we are only concerned with mass ratios, the flavor symmetry breaking scale $\Lambda_f$ remains unresolved at this stage.



This scale can be determined if the physics responsible for new contributions to the dipole operator, which mitigates the observed discrepancy in the muon anomalous magnetic moment, is the very same flavor symmetry \red{ which determines the Yukawa couplings; in this case,} the dipole operator $\bar{L}\sigma_{\mu \nu}P_R\ell F^{\mu \nu}$ \red{will inherit the same basic} flavor structure as the Yukawa interactions. After spontaneous symmetry breaking, it can be expressed as
\begin{align}
    \mc L \supset \frac{e v}{8\pi^2} C_{ij} (\bar{L}_i\sigma_{\mu \nu}P_R \ell_j)F^{\mu \nu} + \text{h.c.} \qquad i,j = 1,2,3, \label{dipoleoperator}
\end{align}
where the dipole matrix $C_{ij}$ in the flavor basis, expressed in units of $\text{GeV}^{-2}$, contains similar factors of $\vev{\varphi_k}/\Lambda_f$. 

Such tree-level Yukawa couplings will receive radiative correction from loops involving the flavons appearing in tree-level diagrams. Because of their similar flavor structures, it is expected that the same loops contributing to the $Y_{ij}$ would also contribute to the $C_{ij}$. Crucially however, each entry in the dipole matrix will in general be multiplied by a different $\mc O(1)$ factor related to the different ways of inserting the external photon line in the dipole operator. This implies that the dipole matrix is not exactly proportional to the Yukawa matrix, and the transformation from the flavor basis to the mass basis will not diagonalize the dipole matrix.\footnote{A similar effect can occur in the soft terms of supersymmetric flavor models generated by a FN mechanism mediated by supergravity, see \cite{Ross:2004qn,Das:2016czs,Lopez-Ibanez:2017xxw,Lopez-Ibanez:2019rgb}.} 

In terms of the dipole matrix, and after rotating to the mass basis, the new physics contributions to the leptonic anomalous magnetic moment $\Delta a_\ell$ is given by
\begin{align}
    \Delta a_\ell = \frac{m_\ell v}{2\pi^2} \text{Re}(C_{\ell \ell}), \quad \ell = e, \mu, \tau,
\end{align}
while the imaginary parts of $C_{\ell\ell}$ are related to the leptonic electric dipole moments (EDMs)\textcolor{black}{\footnote{See Refs.~\cite{Crivellin:2018qmi, Crivellin:2019mvj, Li:2021xmw, Bigaran:2021kmn}, for example, for a combined explanation of $(g-2)_{e,\mu}$ and the relation to EDMs.}}
\begin{align}
    d_\ell  =\frac{ev}{4\pi^2} \text{Im} (C_{\ell \ell}).
\end{align}

Similarly, off-diagonal couplings in the dipole matrix, in the basis of a diagonal $Y_\ell$, contribute to LFV processes, in particular to the radiative decays:
\begin{align}
\frac{{\rm BR}(\ell\to {\ell^\prime} \gamma)}{{\rm BR}(\ell\to \ell^\prime \nu \bar{\nu}^\prime)} = \frac{3\alpha}{\sqrt{2}\pi \,G_F^3\, m_\ell^2} \left( 
|C_{\ell \ell^\prime}|^2 + |C_{\ell^\prime \ell}|^2\right)\,.
\end{align}

\noindent The constraints on the dipole matrix from these processes are summarized in Table \ref{table:DipoleConst}.

\begin{table}[h!]
\centering
{\renewcommand{\arraystretch}{1.3}
\resizebox{0.77\columnwidth}{!}{
\begin{tabular}{c l c}
    \toprule
    \bf Observable & \multicolumn{1}{c}{\bf \boldmath Limit on coefficient} 
    & \bf C.L.\\ 
    \midrule
    $\Delta a_e = \left(4.8 \pm 3.0\right) \times 10^{-13}$ \cite{Morel:2020dww} & ${\rm Re}\left(C_{ee}\right) \approx \left[-0.2,\, 1.7\right] \times 10^{-10}\, {\rm GeV}^{-2}$ & $95\, \%$ \\
    $\Delta a_\mu = \left(251 \pm 59\right) \times 10^{-11}$ \cite{Brown:2001mga,Aoyama:2020ynm,PhysRevLett.126.141801} & ${\rm Re}\left(C_{\mu\mu}\right) \approx \left[1.0,\, 2.8\right] \times 10^{-9}\, {\rm GeV}^{-2}$ & $95\, \%$ \\
    $-0.007 < \Delta a_\tau < 0.005$ \cite{Gonzalez-Sprinberg:2000lzf,Eidelman:2007sb} & ${\rm Re}\left(C_{\tau\tau}\right) \approx \left[-5.9,\, 5.0\right] \times 10^{-4}\, {\rm GeV}^{-2}$ & $95\, \%$ \\
    \midrule
    $d_e < 1.1\times 10^{-29}~ e\, {\rm cm}$ \cite{ACME:2018yjb} & $\left|{\rm Im}\left(C_{ee}\right)\right| \lesssim 9.0\times 10^{-17}\, {\rm GeV}^{-2}$ & $90\, \%$ \\
    $d_\mu < 1.9\times 10^{-19}~ e\, {\rm cm}$ \cite{Muong-2:2008ebm} & $\left|{\rm Im}\left(C_{\mu\mu}\right)\right| \lesssim 1.5\times 10^{-6}\, {\rm GeV}^{-2}$ & $95\, \%$ \\
    $d_\tau < 4.5\times 10^{-17}~ e\, {\rm cm}$ \cite{Belle:2002nla} & $\left|{\rm Im}\left(C_{\tau\tau}\right)\right| \lesssim 3.7\times 10^{-4}\, {\rm GeV}^{-2}$ & $95\, \%$ \\
    \midrule
    ~${\rm BR}\left(\mu\to e \gamma \right)\leq 4.2 \times 10^{-13}$~ \cite{MEG:2016leq} & ~$\left|C_{e\mu}\right|,\, \left|C_{\mu e}\right| \lesssim\, 3.9 \times 10^{-14}\, {\rm GeV}^{-2}$~ & $90\, \%$ \\
    ~${\rm BR}\left(\tau \to e \gamma \right) \leq 3.3 \times 10^{-8}$~ \cite{BaBar:2009hkt}  & ~$\left|C_{e\tau}\right|,\, \left|C_{\tau e}\right| \lesssim\, 4.3 \times 10^{-10}\, {\rm GeV}^{-2}$~ & $90\, \%$ \\
    ~${\rm BR}\left(\tau \to \mu \gamma \right)< 4.2 \times 10^{-8}$~ \cite{Belle:2021ysv} & ~$\left|C_{\mu\tau}\right|,\, \left|C_{\tau\mu}\right| \lesssim\, 5.0 \times 10^{-10}\, {\rm GeV}^{-2}$~ & $90\, \%$ \\
    \bottomrule
\end{tabular}}}
\caption{Current experimental limits on the size of the dipole matrix entries.}
\label{table:DipoleConst}
\end{table}

Following Ref.~\cite{Calibbi:2021qto}, we can assume that  the entries of the dipole matrix are of the order of the corresponding entries of the Yukawa matrix in the flavor basis, $C_{\ell \ell^\prime} ~\gtrsim~ \kappa ~ \frac {Y_{\ell \ell^\prime}}{\Lambda_f^2}$ , 
where we have omitted the different ${\cal O} (1)$ coefficients that, in general, can appear each entry, and $\kappa$ is a global factor that takes care of the relative size of the tree-level Yukawa to the loop contribution to the mass. For example, a typical value of $\kappa \simeq 1/8$ is expected when the dipole is loop-suppressed with respect to the mass, while $\kappa \simeq 2 \pi^2 \simeq 20 $ if the mass has a radiative origin \cite{Calibbi:2021qto}. 

Using the observed discrepancy in the measured muon anomalous magnetic moment to fix the scale $\Lambda_f$, we can then employ the limits on various entries of the dipole operators to obtain the following constraints on the flavor structure of the leptonic Yukawa matrix:

\begin{align}
    Y_\ell\approx y_\tau
    \begin{pmatrix}
              \lambda^{5}& \lesssim\lambda^{8.6}& \lesssim\lambda^{2.8} \\   \lesssim\lambda^{8.6} & \lambda^{2}&\lesssim\lambda^{2.7}\\ \lesssim\lambda^{2.8}& \lesssim\lambda^{2.7}& 1
            \end{pmatrix}, 
    \label{eq:Ye-bounds}
    \end{align}
where $\lambda \simeq 0.225$ is the Wolfenstein parameter.\footnote{\textcolor{black}{The $(11)$ element is determined by requiring the observed mass ratio $m_e/m_\tau \approx \lambda^5$, which could also be accommodated if $Y_\ell^{13} Y_{\ell}^{31} \sim \lambda^5$ while $Y_\ell^{11} \ll \lambda^5$.}} 

In view of the above constraints, our strategy for model building is as follows:
\begin{enumerate}[(i)]
    \item We propose a charged-lepton Yukawa matrix in the flavor basis, arising from the breaking of an underlying flavor symmetry, so that the ratios of its eigenvalues yield the observed charged-lepton mass ratios.
    \item The dipole matrix is then generated with the same flavor structure as the loop corrections to the Yukawas, except for possible differences in the $\mc O(1)$ factors of each entry, determined by the ways of attaching the external photon line in the generating one-loop diagram.
    \item Rotating to the mass basis, we match the contribution of the $(22)$ element of the dipole matrix to the anomalous magnetic moment of the muon, so that the experimentally observed value of $(g-2)_\mu$ is reproduced. From this we determine the scale $\Lambda_f$.
    \item We impose upper bounds from LFV observables to the off-diagonal entries of the dipole matrix in the mass basis, which in turn restricts the structure of the Yukawa matrix.
\end{enumerate}

{\color{black} In general, simultaneously satisfying all of the above constraints in a realistic model is nontrivial, and may not be possible for a low flavor scale $\Lambda_f$.} As a proof of concept that it can be done in principle, in the following two sections we build two explicit models based on the flavor groups $\mc T_{13}$ and $A_5$, respectively.


\section{A $\mc T_{13}$ Model}\label{sec:T13}

$\mc T_{13} \equiv \mc Z_{13} \rtimes \mc Z_{3}$ is an $\mc O(39)$ finite subgroup of $SU(3)$ \cite{bovier1981finite, bovier1981representations, fairbairn1982some,  ishimori2012introduction, Ramond:2020dgm}. It contains two generators $a$ and $b$, related to its $\mc Z_{13}$ and $\mc Z_{3}$ subgroups. These generators are nontrivially related to each other, yielding the presentation
\begin{align*}
    \langle a,b ~|~ a^{13}=b^3=I, bab^{-1}=a^3\rangle.
\end{align*} 
The group has two distinct complex $3$-dimensional representations, a trivial singlet, and a complex singlet. It is the smallest discrete subgroup of $SU(3)$ with two inequivalent complex triplet representations, $\g{3}_1$ and $\g{3}_2$. Each element of the triplets has a unique $\mc Z_{13}$ charge, 
\begin{align}
\begin{gathered}
    \g{3}_1: (\rho^1, \rho^3, \rho^9), \qquad \gb{3}_1: (\rho^{12}, \rho^{10}, \rho^4),\\
    \g{3}_2: (\rho^{2}, \rho^6, \rho^5), \qquad \gb{3}_2: (\rho^{11}, \rho^{7}, \rho^8),
\end{gathered}\label{Z13}
\end{align}
where $\rho^{13} \equiv 1$.
The Kronecker products and Clebsch-Gordan coefficients of the group are summarized in Appendix~\ref{app:T13}. 


An interesting feature of $\mc T_{13}$ is that the Kronecker product of two triplets places the diagonal and off-diagonal terms in different representations. This can be useful in model building, specifically when there are different constraints on the diagonal and off-diagonal elements of the Yukawa matrices. Assuming the SM $SU(2)$ doublets and singlets transform as triplets, while the Higgs transform trivially under $\mc T_{13}$, the Yukawa matrix elements are then generated from the Kronecker product of two lepton triplets. If the two triplet representations are the same, each diagonal and off-diagonal pair can be tagged with a distinct $\mc Z_{13}$ charge, as given by Eq.~\eqref{Z13}.

\red{Although $\mc T_{13}$ has most often found use in describing popular mixing patterns (such as Tribimaximal mixing) in the neutrino sector,\footnote{See Refs.~\cite{ding2011tri, hartmann2011neutrino, Hartmann:2011dn,Perez:2019aqq,CentellesChulia:2019ldn,Perez:2020nqq, Rahat:2020mio, Fong:2021tqj} for the application of $\mc T_{13}$ in BSM model building.} our aim here is to employ it in building a new model capable of producing the flavor structure for the charged-lepton Yukawa matrix given in Eq.~\eqref{eq:Ye-bounds}. We leave an extension of such a model to the neutrino sector for a future work; however, as an example of a model which describes both the charged lepton and neutrino sectors, we present a modified $A_5$ model found in the literature in Section \ref{sec:A5}.}            



\subsection{Tree Level Model}
As a concrete example, we take the $SU(2)$ lepton doublets and singlets to both transform as the $\g{3}_1$ representation of $\mc T_{13}$: $\bar{L} \equiv (\bar{L}_1, \bar{L}_2, \bar{L}_3) \sim \g{3}_1$ and $\ell \equiv (\ell_1, \ell_2, \ell_3) \sim \g{3}_1$. In this basis, the entries $\left(Y_\ell\right)_{ij}$ are given by the coefficients of $\bar{L}_{j}\ell_{i}$. 

Our objective is to build a minimal Yukawa matrix whose diagonalization yields the hierarchical charged-lepton mass ratios. An intuitive understanding of how to generate the desired Yukawa structure can be gained from the fermion bilinears. From the $\mc T_{13}$ Clebsch-Gordan coefficients, we have 
\begin{align}
    \left(
\begin{array}{c}
 \bar{L}_1 \\
 \bar{L}_2 \\
 \bar{L}_3 \\
\end{array}
\right)_{\g{3}_1} \otimes 
    \left(
\begin{array}{c}
 \ell_1 \\
 \ell_2 \\
 \ell_3 \\
\end{array}
\right)_{\g{3}_1} &= 
\left(
\begin{array}{c}
 \bar{L}_1\ell_1 \\
 \bar{L}_2\ell_2 \\
 \bar{L}_3\ell_3 \\
\end{array}
\right)_{\g{3}_2} \oplus
\left(
\begin{array}{c}
 \bar{L}_2\ell_3 \\
 \bar{L}_3\ell_1 \\
 \bar{L}_1\ell_2 \\
\end{array}
\right)_{\gb{3}_1} \oplus
\left(
\begin{array}{c}
 \bar{L}_3\ell_2 \\
 \bar{L}_1\ell_3 \\
 \bar{L}_2\ell_1 \\
\end{array}
\right)_{\gb{3}_1}.
\label{eq:altmod}
\end{align}
As promised, the diagonal and off-diagonal elements of the Yukawa matrix are separated into the $\g{3}_2$ and $\gb{3}_1$ representations, respectively. Each element has a unique $\mc Z_{13}$ charge according to Eq.~\eqref{Z13}, and can therefore be paired with a flavon of conjugate charge.  

For simplicity we take the SM Higgs to be a $\mc T_{13}$ singlet. The Yukawa matrix is then generated by dimension-$5$ and -higher operators of the form $\bar{L}\ell H \varphi$, where $\varphi$ is a (combination of) flavon(s) transforming as a triplet/antitriplet under $\mc T_{13}$.  



The observed hierarchy in the charged-lepton masses suggests a diagonal Yukawa matrix $Y_\ell \sim \text{diag}(\lambda^5, \lambda^2, 1)$. Naively, one could generate this structure by employing three flavons, all transforming as a $\gb{3}_2$, cf. Eq.~\eqref{eq:altmod}, and vacuum values $\lambda^5(1,0,0)$, $\lambda^2(0,1,0)$ and $(0,0,1)$, respectively. In this case however, the hierarchy between the scale of the first and third flavon VEVs would be large, $\mc O(\lambda^5)$. 

Another option is a quasi-diagonal Yukawa structure, where  $Y_\ell^{11}$ is zero and the electron mass is generated from $\mc O(\lambda^{5/2})$ entries in the $Y_\ell^{13-31}$ elements. This can be achieved with a single flavon VEV $\lambda^{5/2}(0,1,0)$ transforming as a $\g{3}_1$, cf. Eq.~\eqref{eq:altmod}, thanks to the separation of diagonals and off-diagonals terms. 

We also require a sizable loop correction to $Y^{22}_\ell$, as a similar diagram will generate the contribution to $(g-2)_{\mu}$ in the dipole matrix. If $Y_\ell^{22}$ is generated by the product of two identical flavons $\varphi_{22}$, one could expect a loop correction from the quartic coupling $(\varphi_{22}\varphi_{22}^*)^2$. {\color{black} Here we make use of the unique $\mc Z_{13}$ charges of the triplets to introduce a useful subscript notation for the flavons, distinguishing them by the elements of $Y_\ell$ to which they couple in vacuum.} 

However, as will be shown, the contribution to the dipole matrix is a product of a negative loop factor and the quartic flavon coupling. Since the required contribution to muon $g-2$ is positive, the flavon quartic coupling must be negative. In this case, one should worry about the stability of the potential; a simple remedy is to include a negative mixed quartic coupling $\beta_{2a} (\varphi_{a}\varphi_{22}^*)^2$ while keeping the couplings $\beta_2 (\varphi_{22}\varphi_{22}^*)^2$ and  $\beta_a (\varphi_{a}\varphi_{a}^*)^2$ such that $\beta_{2a} + \sqrt{\beta_2 \beta_a}$ is positive \cite{Deshpande:1977rw,Klimenko:1984qx,Nie:1998yn,Maniatis:2006fs,Bhattacharyya:2015nca}. $\mc T_{13}$ constrains this flavon $\varphi_a$ to be of the same representation as $\varphi_{22}$, so that the mixed quartic loop contributes to $Y_\ell^{22}$. This implies that $\varphi_a$ must be $\varphi_{33}$, the flavon which couples to $\bar{L}_{3}\ell_{3}$, and therefore that $Y_{\ell}^{33}$ be generated by the product $\varphi_{33}^2$ at tree-level. 

Given the above reasoning, a minimal Yukawa structure emerges where the matrix elements $Y_\ell^{13-31}$, $Y_\ell^{33}$, and $Y_\ell^{22}$ 
are generated from the following Lagrangian
\begin{align} \label{T13Yukawa}
    \mc L_Y^{e} =  \bar{L}\ell H \left[ \frac{1}{\Lambda_f}\varphi_{13} + \frac{1}{\Lambda_f^2}\varphi_{22}\varphi_{22} + \frac{1}{\Lambda_f^2} \varphi_{33} \varphi_{33} \right].
\end{align}
The flavons transform under $\mc T_{13}$ as $\varphi_{33} \sim \gb{3}_1$, $\varphi_{22} \sim \gb{3}_1$, $\varphi_{13} \sim \g{3}_1$, aligning in the flavor vacuum along 
\begin{align}
    \vev{\varphi_{33}} = \epsilon\, (0,0,1)\, \Lambda_f, \quad \vev{\varphi_{22}} = \epsilon^2\, (0,1,0)\, \Lambda_f, \quad \vev{\varphi_{13}} = \epsilon^{9/2}\, (0,1,0)\, \Lambda_f,
\end{align}
where $\epsilon \sim \mc O(\lambda)$, fixed by the observed lepton mass ratios. Notice that we have to add a factor $\epsilon$ in the VEV of $\varphi_{33}$, so that the mass of $\tau$ is reproduced with $v /\sqrt{2}= 174$~GeV with an ${\cal O}(1)$ coefficient. Furthermore, in this way, higher order operators like $\bar{L}\ell H\varphi_{a}(\varphi_{33}^* \varphi_{33})^n$ are suppressed by a factor $\epsilon^{2n}$ with respect to the leading order contribution. 

\begin{figure}[t]
\centering
\subfloat[$Y^{13-31}_\ell$\label{fig:Y13}]{
\includegraphics[scale=1]{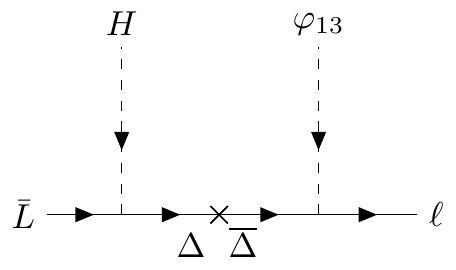}  
}\\
\subfloat[$Y^{22}_\ell$\label{fig:Y22}]{
\includegraphics[scale=1]{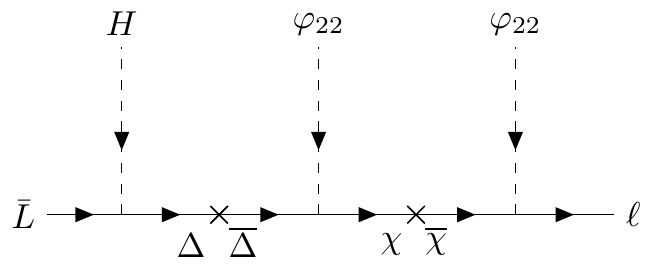}   
}
\subfloat[$Y^{33}_\ell$\label{fig:Y33}]{
\includegraphics[scale=1]{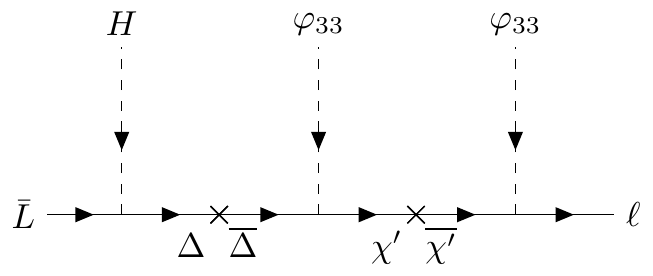}   
} \\
\subfloat[cubic\label{fig:cubic}]{
\includegraphics[scale=1]{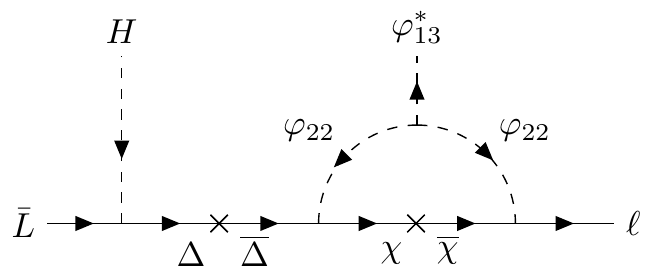} 
}
\subfloat[quartic\label{fig:quartic}]{
\includegraphics[scale=1]{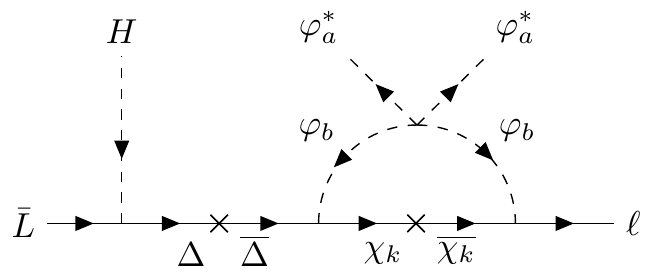} 
} 


\label{fig:FNdiag}
\caption{Feynman diagrams for effective operators generating the charged lepton Yukawa matrix. Here $\times$ denotes a mass insertion. Figs.~\ref{fig:Y13}-\ref{fig:Y33} show the tree-level contributions while Figs.~\ref{fig:cubic}-\ref{fig:quartic} show loop corrections. For Fig.~\ref{fig:quartic}, the mediator $\chi_k \equiv \chi (\chi')$ when $\varphi_{b} \equiv \varphi_{22} (\varphi_{33})$.}
\end{figure}

At tree level, the operators in Eq.~\eqref{T13Yukawa} can be generated using heavy vectorlike messengers as mediators, as shown in the Feynman diagrams of Figs.~\ref{fig:Y13}-\ref{fig:Y33}. \textcolor{black}{In principle, without specifying the full UV theory, one would naively expect mediators of all possible representations, generating additional vertices. The simplest way to forbid such dangerous operators is then to restrict the messenger spectrum, which we do here by including only the} {\color{black} three mediators $\Delta$, $\chi$ and $\chi'$. This is the minimal set of mediators required to generate the operators in Eq.~\eqref{T13Yukawa}. The mediators are $SU(2)$ singlets, and their $\mc T_{13}$ charges, along with the charges of the other fields in the model, are listed in Table~\ref{table:model2}. 

We note that one could also generate the same effective operators by coupling the Higgs to $\ell$ instead of $\bar{L}$ in the relevant diagrams, in which case some of the mediators would be $SU(2)$ doublets. The choice in Figs.~\ref{fig:Y13}-\ref{fig:Y33} is motivated by the relatively relaxed bounds on singlet mediator masses compared to doublets \cite{Bissmann:2020lge}. As we will see later, successfully reproducing the correct value for $(g-2)_{\mu}$ requires a low flavor scale $\Lambda_f$ in this model, which might be in tension with LHC bounds on $SU(2)$ doublet mediators \cite{Bissmann:2020lge}. }

While limiting the allowed mediators restricts the possible vertices to a great extent, dangerous terms allowed by $\mc T_{13}$ still remain. To further protect the desired flavor structure of $Y_{\ell}$ against unwanted contributions from these vertices, we introduce an abelian $\mc Z_{n}$ ``shaping symmetry''; {\color{black} the origin of this shaping symmetry and the restricted mediator spectrum must then be addressed in a UV completion of the model. }

\begin{table}[t]\centering
\renewcommand\arraystretch{1.1}
\begin{tabularx}{\textwidth}{@{}l  Y Y  Y  Y Y Y  Y Y Y @{}}
    \toprule
    Fields &  $\bar{L}$ & $\ell$ & $H$ & $\varphi_{33}$ & $\varphi_{22}$ & $\varphi_{13}$ & $\Delta$ & $\chi$ & $\chi'$ \\
    \midrule
    $SU(2)_L$ & $\g{2}$ & $\g{1}$ & $\g{2}$ & $\g{1}$ & $\g{1}$ & $\g{1}$ & $\g{1}$ & $\g{1}$ & $\g{1}$ \\
    $\mc T_{13}$ & $\g{3}_1$ & $\g{3}_1$ & $\g{1}$ & $\gb{3}_1$ & $\gb{3}_1$ & $\g{3}_1$ & $\gb{3}_1$ & $\g{1}$ & $\g{1}$ \\
    $\mc Z_{4}$ & $\g{\eta^{1}}$ &  $\g{1}$ &  $\g{\eta^{1}}$ &  $\g{\eta^{1}}$ &  $\g{\eta^{3}}$ &  $\g{\eta^{2}}$ &  $\g{\eta^{2}}$ &  $\g{\eta^{3}}$ &  $\g{\eta^{1}}$\\
    \bottomrule
\end{tabularx} 
\caption{Transformation properties of matter, scalar and messenger fields. Here  $\eta^{4} = 1$. The $\mc Z_{4}$ `shaping' symmetry prevents unwanted tree-level operators.
}
\label{table:model2}
\end{table}


\subsection{``Shaping'' Symmetry}
There are six ``expected" vertices involving two fermions and a scalar in this model, as seen from Figs.~\ref{fig:Y13}-\ref{fig:Y33}: $\bar{L}\Delta H$, $\ell \overline{\Delta} \varphi_{13}$, $\ell \overline{\chi} \varphi_{22}$, $\ell \overline{\chi'} \varphi_{33}$, $\chi \overline{\Delta}\varphi_{22}$ and $\chi' \overline{\Delta} \varphi_{33}$. On the other hand, ten ``unwanted'' vertices are still allowed by the $\mc T_{13}$ and SM charges, listed below in two categories
\begin{align}
    \mathrm{(i)}\quad
    &\ell \overline{\Delta} \varphi_{33}^*,\ \ell \overline{\Delta} \varphi_{22}^*,\ \ell \overline{\chi} \varphi_{13}^*,\ \ell \overline{\chi'} \varphi_{13}^*,\ \chi \overline{\Delta} \varphi_{13}^*,\ \chi' \overline{\Delta} \varphi_{13}^*, \\
    \mathrm{(ii)}\quad
    &\ell \overline{\chi} \varphi_{33},\ \ell \overline{\chi'} \varphi_{22},\ \chi\overline{\Delta}\varphi_{33},\ \chi'\overline{\Delta}\varphi_{22}.
\end{align}
These vertices can be prevented by introducing a $\mc Z_n$ ``shaping'' symmetry. As there are nine fields and six vertices in the model, at least three of the fields will have independent $\mc Z_n$ charges. 

Denoting these charges with a $[\ \cdot\ ]$ notation, suppose $[\bar{L}] = x$, $[\ell] = y$ and $[H] = z$. Then from the desired vertices, one has that
\begin{equation}
\begin{gathered}
    [\Delta] [\Delta]= -x-z,\quad  [\varphi_{13}] = -x-y-z, \quad [\varphi_{22}] =  -\frac{x+y+z}{2}, \\ [\varphi_{33}] = \frac{n}{2}-\frac{x+y+z}{2},\ [\chi] =  \frac{-x+y-z}{2},\ [\chi'] = \frac{-x+y-z}{2} - \frac{n}{2}, \label{Zncharge}
\end{gathered}
\end{equation}
modulo $n$. Note that the quartic scalar vertex $(\varphi_{33}  \varphi_{22}^*)^2 $ required by the model is always allowed. The charges of $\varphi_{22}$ and $\varphi_{33}$ (and the corresponding mediators $\chi$ and $\chi'$) have been separated by $n/2$ so that the otherwise allowed operator $\bar{L}\ell H \varphi_{22} \varphi_{33}$, which contributes to $Y_\ell^{23-32}$ at $\mc O(\epsilon^3)$, is no longer permitted.

The remaining dangerous operators in category (ii) are prevented by any $\mc Z_{n}$ symmetry for the charge assignment given by Eq.~\eqref{Zncharge}. This follows from the fact that operators in the first category are not allowed when 
\begin{align}
    3(x+y+z) &\neq 0\ \text{mod } n \label{condition1}.
\end{align}
On the other hand, from Eq.~\eqref{Zncharge}, requiring all the field charges to be integers, we have
\begin{align}
    x+y+z &= 0\ \text{mod } 2, \label{condition2}\\
    y-z-x &= 0\ \text{mod } 2,\label{condition3}\\
    \text{and}\quad n &= 0\ \text{mod } 2.\label{condition4}
\end{align}
The case for $n=2$ is ruled out as then Eqs.~\eqref{condition1} and \eqref{condition2} would be in conflict. The next case $n=4$ is viable, for example, with the charges shown in Table \ref{table:model2} using $y=0$, $x=z=1$.


\subsection{Loop Corrections}
With these $\mc Z_{n}$ charges, the flavon potential is restricted to have the following form
\begin{align}
    V_\varphi & = -\mu_1^2\, \varphi_{13}^* \varphi_{13} \;-\; 
            \mu_2^2\, \varphi_{22}^* \varphi_{22} \;-\; 
            \mu_3^2\, \varphi_{33}^* \varphi_{33} \;+\; 
            \left( A\, \varphi_{13}^*\varphi_{22}^2 + {\rm h.c.} \right) \;+\; \frac{\beta_1}{2}\, (\varphi_{13}^*\varphi_{13})^2   \nonumber \\
            & + \frac{\beta_2}{2}\, (\varphi_{22}^*\varphi_{22})^2 \;+\;
            \frac{\beta_3}{2}\, (\varphi_{33}^*\varphi_{33})^2 \;+\;
            \left[\, \frac{\beta_4}{2}\, (\varphi_{22}^*\varphi_{33})^2 + {\rm h.c.} \,\right] + \dots, \label{eq:Vfit} 
\end{align}
where the dots stand for additional operators related to all possible contractions of quartic couplings of the form $\varphi_a^* \varphi_a \varphi_b^* \varphi_b$; the $\mu_i$ are the masses of the flavons, pressumably $\mc O(\Lambda_f)$, $A\sim {\cal O}(\Lambda_f)$ is a cubic coupling and the $\beta_i$ are quartic couplings.

The cubic and quartic couplings present in the potential induce loop corrections to the Yukawas. In Fig.~\ref{fig:cubic}, the VEV of $\varphi_{13}^*$ contributes to  $Y_\ell^{13-31}$, similar to the diagram in Fig.~\ref{fig:Y13}. The conjugate of the cubic term yields another diagram like Fig.~\ref{fig:Y13}, with  $(\varphi_{22}^*\varphi_{22}^*)$ coupling to the $\varphi_{13}$. Fortunately however, the bilinear $(\varphi_{22}^*\varphi_{22}^*)$ in the representation of $\varphi_{13}^*$ is zero in the vacuum, and does not contribute to $Y_{\ell}^{22}$. 

The quartic interactions yield loop corrections through Feynman diagrams of the form of Fig.~\ref{fig:quartic}. The pure quartic terms $(\varphi_{a} \varphi_{a}^*)^2$ contribute to $Y_\ell^{a}$. Since there is no tree-level diagram with two $\varphi_{13}$'s coupling to fermions, $Y_{\ell}^{13-31}$ does not receive any loop correction from the pure quartic term. On the other hand, the mixed quartic terms $(\varphi_{22}^*\varphi_{33})^2$  and $(\varphi_{33}^*\varphi_{22})^2$ contribute to $Y_\ell^{22}$  and $Y_\ell^{33}$, respectively.


\subsection{Predictions}
The Yukawa matrix generated by the operators of Eq.~\eqref{T13Yukawa} can be expressed as
\begin{align} \label{eq:YFNfit}
    Y_\ell = \epsilon^2 \left( 
    \begin{array}{ccc}
            0 & 0 & \epsilon^{5/2} \\
            0 & \epsilon^2 & 0 \\
            \epsilon^{5/2} & 0 & y_1 \\
    \end{array}\right), 
\end{align}
where we have introduced an $\mc O(1)$ coefficient $y_1$ to account for a slightly different magnitude in the VEV of $\varphi_{33}$ with respect to the others.
From Eq.~\eqref{eq:YFNfit}, the following mass ratios are obtained
\begin{eqnarray}
    \frac{m_e}{m_\tau} & = & 1+\frac{y_1^2}{2\, \epsilon^5}\, \left(1-\sqrt{1+\frac{4\epsilon^5}{y_1^2}}\right) ~\simeq~ \frac{\epsilon^5}{y_1^2} ~\sim~ 0.0003, \\ 
    \frac{m_\mu}{m_\tau} & = & \frac{2\, \epsilon^2}{y_1}\frac{1}{1+\sqrt{1+\frac{4\epsilon^5}{y_1^2}}} 
    ~\simeq~ \frac{\epsilon^2}{y_1} ~\sim~0.045,
\end{eqnarray}
which, to a first approximation, are consistent with observation for $\epsilon\sim 0.15$ and $y_1\sim 0.5$.

Radiative corrections to the Yukawa matrix of Eq.~\eqref{eq:YFNfit} are generated by the diagrams in Fig. \ref{fig:cubic}-\ref{fig:quartic}, where the flavon lines can be closed through a cubic coupling $A$ or quartic coupling $\beta_i$.
They yield 
\begin{align} \label{eq:Yradfit}
    \delta Y_\ell = \epsilon^2\, \frac{f_1(x_\varphi^2)}{32\, \pi^2}\,  \left( \begin{array}{ccc}
            0 & 0 & 2\, y_2\, y_4\, \epsilon^{5/2} \\
            0 & \left(\beta_2\, y_2 + \beta_4\, y_3 \right) \epsilon^2 & 0 \\
            2\, y_2\, y_4\, \epsilon^{5/2} & 0 & y_1\, \left(\beta_3\, y_3 + \beta_4\, y_2 \right) \end{array}\right),
\end{align}
where $f_1(x)$ is the loop function given by
\begin{align} \label{eq:f1}
    f_1(x) = \frac{1+\log x-x}{\left(1-x\right)^{2}}<0,
\end{align}
and $x_\varphi\equiv m_\varphi/M_\chi$, with $m_\varphi$ and $M_\chi$ generic masses for the flavons and heavy mediators.

In principle, the radiative correction for each element has a different loop factor $f_1(m_{\varphi_a}^2/M_\chi^2)$, where $m_{\varphi_a}$ corresponds to the mass of the flavon contributing to the particular element. If $m_{\varphi_a} \sim \vev{\varphi_a}$, the masses of these flavons are expected to differ by some orders of magnitude. However, within the loop function, this variation can effectively be approximated as a change in the ${\cal O}$(1) coefficient. We therefore take the loop factor as common to all elements, introducing new $\mc O(1)$ coefficients $y_2$ and $y_3$ to take into account the possible differences in mass of $\varphi_{22}$ and $\varphi_{33}$. Similarly, the cubic coupling $A$ can be absorbed in the $\mc O(1)$ factor $y_4\equiv A/M_\chi$.

\begin{figure}[t]
\centering
\subfloat[cubic\label{fig:Ccubic}]{
\includegraphics[scale=1]{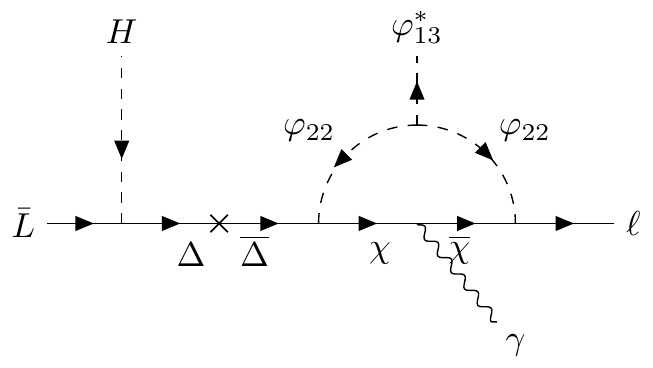}
}
\subfloat[quartic\label{fig:Cquartic}]{
\includegraphics[scale=1]{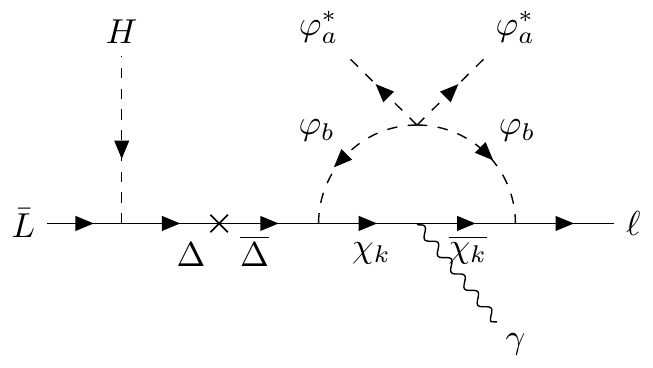}
}
\caption{Diagrams contributing to the dipole operators. 
}
\label{fig:dipolediag}
\end{figure}

The dipole operator is generated through diagrams similar to those that give radiative correction to the Yukawa couplings, albeit with a photon radiated from the charged particle inside the loop, see Fig. \ref{fig:dipolediag}.
The resulting matrix therefore has the same basic structure as the radiative correction to the Yukawas in Eq.~\eqref{eq:Yradfit}.
However, as it is a dimension-6 operator, an explicit dependence on the flavor scale $\Lambda_f$ is present\footnote{We assume here that this scale, where the flavor dynamics act, is roughly the same as the mediator masses.}: 
\begin{align} \label{eq:Cfit}
    C_\ell = \frac{f_2(x_\varphi^2)}{16}\, \frac{\epsilon^2}{\Lambda_f^2} \,\left( \begin{array}{ccc}
            0 & 0 & 2\, y_2\, y_4\, \epsilon^{5/2} \\
            0 & \left(\beta_2\, y_2 + \beta_4\, y_3 \right) \epsilon^2 & 0 \\
            2\, y_2\, y_4\, \epsilon^{5/2} & 0 & y_1\, \left(\beta_3\, y_3 + \beta_4\, y_2 \right) \end{array}\right),
\end{align}
where again we have taken the loop function $f_2(x)$, given by
\begin{align} \label{eq:f2}
    f_2(x) = -\frac{1+4 x\left(1+\log x\right)-x^2\left(5-2 \log x\right)}{2\left(1-x\right)^{4}}<0.
\end{align}
as common to all elements while capturing the differences in the flavon masses through the factors $y_2$ and $y_3$.

\begin{figure}
    \centering
    \includegraphics[width=0.48\textwidth]{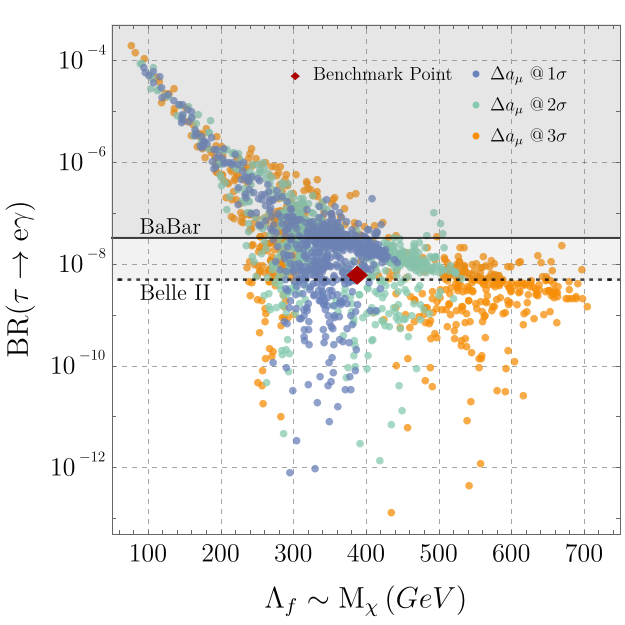}
    \hspace{0.35cm}
    \includegraphics[width=0.48\textwidth]{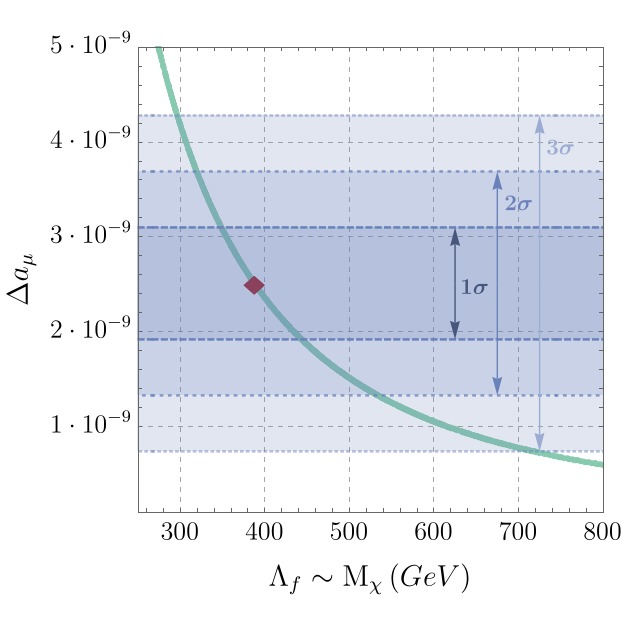}
    \caption{Branching ratio of the LFV decay $\tau \to e\gamma$ vs. mass of the heavy mediator ($\sim \Lambda_f$) for several points that correctly reproduce the mass of the charged leptons and $\Delta a_\mu$.
    Our benchmark point in Table~\ref{table:fit} corresponds to the red diamond.
    }
    \label{fig:T13fits}
\end{figure}

In total, nine free parameters determine the charged-lepton masses and anomalous magnetic moments in this model.
To find their best-fit values, we scan and minimize over the $\chi^2$-function given in Eq.~\eqref{appeq:X2}.
The fine tuning of the model is also computed for each fit, as indicated in Eq.~\eqref{eq:FTfunct}.
In the left panel of Fig. \ref{fig:T13fits}, we display a set of points with fine tuning less than $10$ \cite{Fedele:2020fvh} which provide good fits to both the masses and $(g-2)_{\mu}$. Mediator masses of up to $430$ GeV ($705$ GeV) can accommodate the measured discrepancy in the anomalous magnetic of the muon at the $1\sigma$ ($3\sigma$) level.
We select as a benchmark point the one with the heaviest mediator mass which exactly reproduces the central value of $(g-2)_{\mu}$, see Table~\ref{table:fit}.

\begin{table}[ht]
\centering
\renewcommand\arraystretch{1.1}
\begin{tabularx}{0.7\textwidth}{c c c c c c c c c}
    \toprule
    $y_1$ & $y_2$ & $y_3$ & $y_4$ & $\beta_2$ & $\beta_3$ & $\beta_4$ & $\epsilon$ & $x_\varphi$ \\
    \midrule
    $0.50$ & $0.50$ & $3.50$ & $0.70$ & $0.25$ & $1.56$ & $-6.24$ & $0.15$ & $0.08$ \\
    \bottomrule
    \end{tabularx} 
\caption{Benchmark point for the $\mc T_{13}$ model.}
\label{table:fit}
\end{table}

As detailed in Table~\ref{table:pred}, our representative point in Table~\ref{table:fit} successfully reproduces both the charged lepton masses and the anomalous magnetic moment of the muon.
Our model contribution to the electron $g-2$ remains small, so that the total prediction (SM+flavor model) and the experimental result in Ref. \cite{Morel:2020dww} remain compatible at $1.6\,\sigma$.
Regarding LFV, while the transitions $\mu\to e$ and $\tau\to\mu$ are absent in this model, the flavor-changing decay $\tau \to e \gamma$ has a sizable branching ratio which is predicted to be below present limits but still testable with future sensitivity \cite{Aushev:2010bq} (see Fig. \ref{fig:T13fits}, left).

\begin{table}[ht]
\centering
\renewcommand\arraystretch{1.1}
\begin{tabularx}{0.65\textwidth}{c c c c}
    \toprule
    $m_e$ (MeV) & $m_\mu$ (GeV) & $m_\tau$ (GeV) & $\Delta a_\mu$ \\
    \midrule
    $0.507$ & $0.103$ & $1.806$ & $2.51\times 10^{-9}$ \\
    \toprule
    $\Delta a_e$ & $\Delta a_\tau$ & {\small ${\rm BR}(\tau \to e \gamma)$} & {\small $M_\chi$ (GeV)} \\
    \midrule
    $3.64\times 10^{-15}$  & $-1.07\times 10^{-7}$ & $6.52\times 10^{-9}$ & $388$ \\
    \bottomrule
\end{tabularx} 
\caption{Predictions at the benchmark point of the $\mc T_{13}$ model.}
\label{table:pred}
\end{table}

Finally, the right panel of Fig. \ref{fig:T13fits} displays how the model prediction for the anomalous magnetic moment of the muon varies when the general mass for the heavy mediators is modified for our chosen benchmark point.
In this case, we scan around the parameter values in Table~\ref{table:fit} and select sets which correctly reproduce the charged-lepton masses.
We note that for this specific point in the parameter space, a maximum mass of $710$ GeV can be reached if the prediction for the muon $g-2$ is allowed to fall within the $3\sigma$ range.

The relevant collider bounds on the masses of the mediators in our model come from ATLAS searches for vector-like leptons (VLLs) \cite{ATLAS:2015qoy}, which exclude $SU(2)$ singlets with masses in the range $114$-$176$ GeV, and CMS searches for $SU(2)$ doublets coupling only to taus \cite{CMS:2019hsm}, which rule out masses between $120-790$ GeV.
We note however that these limits have been obtained assuming simplified models.
A more detailed discussion can be found in Ref.~\cite{Bissmann:2020lge}, where the authors interpret the experimental findings in the context of flavorful VLLs.
They conclude that the masses of these particles have to be above $300$ GeV ($800$ GeV), if they transform as singlets (doublets) of $SU(2)$.
These limits are satisfied in our model, where the heavy mediators are defined as $SU(2)$ singlets.
If they had instead transformed as doublets, there would be tension with these bounds, as the largest possible mass capable of reproducing $(g-2)_\mu$ in our model is around $700$ GeV.

\section{An $A_5$ model} \label{sec:A5}
$A_5$ is the nonabelian discrete group composed of the even permutations on five objects. It has $60$ elements and five irreducible representations (irrep): one singlet $\g{1}$, two triplets $\g{3}$ and $\g{3}^\prime$, one tetraplet $\g{4}$ and one pentaplet $\g{5}$ ($1^2 + 2\times 3^2 + 4^2 + 5^2 = 60$). It can be generated by two elements, $s$ and $t$, and is given by the presentation \cite{Luhn:2007yr}
\begin{align}
    \langle s,t\ |\ s^2 \,=\, (s\, t)^3 \,=\, t^5 \,=\, I  \rangle.
\end{align}
The Kronecker products and Clebsch-Gordan coefficients of $A_5$ are given in Appendix \ref{app:A5}. 

$A_5$ has been used extensively as a flavor symmetry for explaining the observed lepton masses and mixings.\footnote{Some early examples include Refs.~\cite{Everett:2008et,Chen:2010ty,Feruglio:2011qq,Ding:2011cm}.} Typically, the flavor symmetry is broken down into different residual symmetries in the charged lepton and in the neutrino sector. These residual symmetries constrain the form of the flavon VEVs in each sector.

Following the analysis of Refs.~\cite{Li:2015jxa, DiIura:2015kfa, Ballett:2015wia}, we consider {$\mathcal{G}_e={\mc Z}_5$} as the residual symmetry in the charged-lepton sector. This implies that, in vacuum, the flavons $\varphi_\g{i}^e$ must be symmetric under the transformations of the generators $Q_\g{i}$ corresponding to the representation $\g{i}$ of $A_5$ \cite{Ding:2011cm, DiIura:2018fnk}:
\begin{align} \label{eqn:feCond} 
    Q_\g{i}\, \langle \varphi^e_\g{i} \rangle = \langle \varphi^e_\g{i} \rangle. 
\end{align}
Under this condition, non-zero VEVs are possible only for the triplet and pentaplet representations, and their vacuum alignments are of the form
\begin{equation} \label{eqn:chargedVEV}
        \langle \varphi^e_{\bf 3}\rangle = \left(\epsilon_3,\, 0,\, 0 \right)\, \Lambda_f,\qquad
        \langle \varphi^e_{\bf 3'} \rangle = \left(\epsilon_{3'},\, 0,\, 0 \right)\, \Lambda_f,\qquad
        \langle \varphi^e_{\bf 5} \rangle = \left(\epsilon_{5},\, 0,\, 0,\, 0,\, 0\right)\, \Lambda_f,
\end{equation}
where $\epsilon_3,\, \epsilon_{3'}$ and $\epsilon_5$ are dimensionless real parameters. On the other hand, following Ref.~\cite{DiIura:2018fnk}, the neutrino sector respects a different residual symmetry, $\mathcal{G}_\nu={\mc Z}_2\times {\rm CP}$, and the VEVs invariant under this symmetry are,
\begin{align} \label{eqn:nuVEV}
        \langle \varphi^\nu_{\bf 4} \rangle = \left(\begin{array}{c} w_r - i\, w_i \\ (1 + 2 \varphi) w_r - i\, w_i \\ (1 + 2 \varphi) w_r + i~ w_i\\w_r + i~ w_i
        \end{array}\right),\quad
        \langle \varphi^\nu_{\bf 5} \rangle = \left(\begin{array}{c}  \sqrt{\frac{2}{3}} ( z_{r1} + z_{r2})\\ - z_{r1} + i ~\varphi\,z_{i} \\   z_{r2} - i \,z_{i} \\   z_{r2} + i\, z_{i}
         \\ z_{r1} + i ~\varphi z_{i}  \end{array}\right),
\end{align}
where $w_r, w_i, z_{r1}, z_{r2}$ and $z_i$ are real parameters of mass dimension one. An appropriate $\mc Z_{n}$ ``shaping'' symmetry then keeps the flavons $\varphi^{e}$ and $\varphi^{\nu}$ restricted to the charged lepton and the neutrino sectors, respectively. 

\subsection{Tree Level Model}
\begin{figure}[t]
\centering
\subfloat[\label{fig:one}]{
\includegraphics[scale=1]{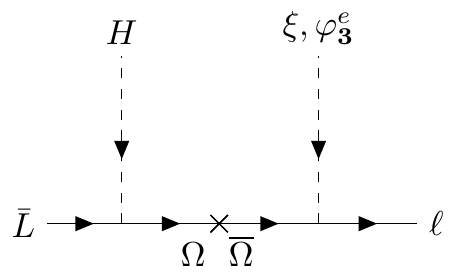}  
}
\subfloat[\label{fig:two}]{
\includegraphics[scale=1]{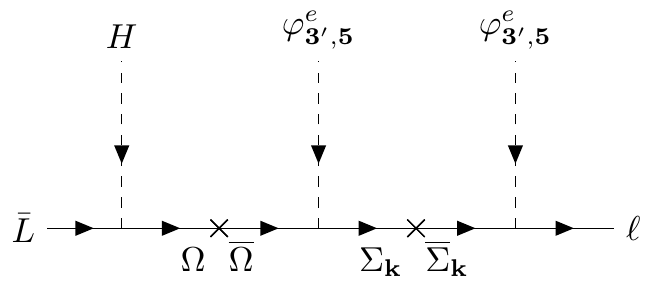}  
}
\\
\subfloat[\label{fig:loop}]{
\includegraphics[scale=1]{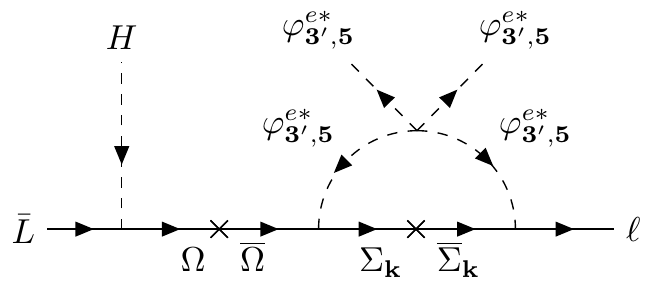}
} 
\caption{Tree level and radiative contributions to the diagonal charged-lepton Yukawa matrix in the $A_5$ model. }
    \label{fig:A5Y}
\end{figure}

Following Ref.~\cite{Ding:2011cm}, we assign the leptons to triplets under $A_5$: $\bar{L} \equiv (\bar{L}_1, \bar{L}_2, \bar{L}_3) \sim \g{3}$, and $\ell \equiv (\ell_1, \ell_3, \ell_2) \sim \g{3}$, where the ordering of the components of the fields has been chosen to be different in order to ensure a completely diagonal charged lepton Yukawa matrix is generated. Four flavons $\xi \sim \g{1}$, $\varphi_{\g{3}}^e \sim \g{3}$, $\varphi_{\g{3'}}^e \sim \g{3'}$ and $\varphi_{\g{5}}^e \sim \g{5}$ participate in the charged-lepton Yukawa interactions. Assigning the $\mc Z_{n}$ charges of the fields as  
\begin{align} \label{ZnChargedLepton}
    [{\bar{L}} \ell H] = q,\quad [\xi] = [\varphi^e_{\bf 3}] = n- q,\quad [\varphi^e_{\bf 5}] = [\varphi^e_{\bf 3'}] = (n-q)/2,
\end{align}
the charged-lepton Yukawas arise from the Lagrangian 
\begin{align}\label{eqn:CharSuperPot}
    {\cal L}^e_{Y} \;=\;  {\bar L}\, \ell\, H \, \left[ \frac{1}{\Lambda_f}(\xi + \varphi^e_{\bf 3}) + \frac{1}{\Lambda^2_f} \big(\varphi^e_{\bf 3'} \varphi^e_{\bf 3'} + \varphi^e_{\bf 5} \varphi^e_{\bf 5}\big)\right],
\end{align}
{\color{black} At tree level, these effective operators can be realized by the renormalizable vertices of the Feynman diagrams in Fig.~\ref{fig:one} and \ref{fig:two}.} 

Here the flavor interactions are mediated by the vectorlike fields $\Omega$ and $\Sigma_{\g{k}}$. Both of these fields are chosen to be $SU(2)$ singlets to evade the collider bounds, similar to the $\mc T_{13}$ model. Under $A_5$, $\Omega$ transforms as a $\g{3}$, whereas $\Sigma_{\g{k}}$ transforms as either a $\g{4}$ or $\g{5}$ (either  $\g{3}, \g{3'}, \g{4}$ or $\g{5}$) if it couples to $\varphi_{\g{3'}}^e$ ($\varphi_{\g{5}}^e$). 

These diagrams realize the following contractions
\begin{align*}
    \left((\bar{L} H)_\g{3} (\ell \xi)_{\g{3}}\right)_\g{1}, \quad \left((\bar{L} H)_\g{3} (\ell \varphi_{\g{3}}^e)_{\g{3}}\right)_\g{1}, \quad
    \left(\left((\bar{L} H)_\g{3} \varphi_{\g{3'}} \right)_\g{j} (\ell \varphi_{\g{3'}})_{\g{j}}\right)_\g{1}, \quad  \left(\left((\bar{L} H)_\g{3} \varphi_{\g{5}} \right)_\g{k} (\ell \varphi_{\g{5}})_{\g{k}}\right)_\g{1},
\end{align*}
where $\g{j} = \g{4}, \g{5}$ and $\g{k} = \g{3}, \g{3'}, \g{4}, \g{5}$.
At tree level, the Lagrangian in Eq.~\eqref{eqn:CharSuperPot} yields the diagonal Yukawa matrix
\begin{align}
    Y_\ell = \left(
\begin{array}{ccc}
 \epsilon_\xi + 3\,\epsilon^2_{3'} + 7\,\epsilon^2_{5} & 0 & 0 \\
 0 & \epsilon_\xi + \epsilon_3 + 3\, \epsilon^2_{3'} - 8\, \epsilon^2_5  & 0 \\
 0 & 0 & \epsilon_\xi - \epsilon_3 + 3\, \epsilon^2_{3'} - 8\, \epsilon^2_5
\end{array}
\right), \label{eq:Yl3exact}
\end{align}
utilizing a similar $\epsilon$ notation as for the $\mc T_{13}$ model, and where the different $\mc O(1)$ numbers correspond to different ways of contracting the operators in Eq.~\eqref{eqn:CharSuperPot}.

\subsection{Loop Corrections}
{\color{black} Radiative corrections are generated by quartic flavon couplings of the form} $\beta_{i} (\varphi_{\g{i}} \varphi_{\g{i}}^*)^2$, where $\g{i} = \g{3'}, \g{5}$, as shown in the Feynman diagram of Fig.~\ref{fig:loop}. These loop contributions are given by
\begin{align}
\delta Y_\ell = \frac{f_1(x_\varphi)}{32\pi^2}
\left(\begin{array}{ccc}
3\, \epsilon_{3'}^2\, \beta_{3'}\, y_3 + 7\, \epsilon_5^2\, \beta_5\, y_5 &  0  &  0 \\
0 & 3\, \epsilon_{3'}^2\, \beta_{3'}\, y_3 - 8\, \epsilon_5^2\, \beta_5\, y_5  &  0 \\
0 &  0  & 3\, \epsilon_{3'}^2\, \beta_{3'}\, y_3 - 8\, \epsilon_5^2\, \beta_5\, y_5 \\
\end{array}
\right), \label{eq:dYl3exact}
\end{align}
where 
$f_1(x)$ is the loop function in Eq.~\eqref{eq:f1}.
As in the ${\cal T}_{13}$ model, $y_3$ and $y_5$ are ${\cal O}(1)$ coefficients that account for possible differences in mass of $\varphi_{3'}$ and $\varphi_5$.
The values of $\epsilon_3$, $\epsilon_{3'}$ and $\epsilon_5$ are fixed by the conditions of reproducing the correct charged lepton masses:
\begin{align} \label{eq:ycondA5}
    \epsilon_3 &= \frac{1}{\sqrt{2}\, v}\left( m_{\mu} - m_{\tau} \right) \simeq -0.0048, \nonumber \\
    \epsilon_5^2 &\simeq \frac{\sqrt{2}}{15\, v} \left( m_e - \frac{m_{\mu} - m_\tau}{2} \right) \simeq -0.00036, \\
    \epsilon_\xi + 3\, \epsilon_{3'}^2 & \simeq \frac{\sqrt{2}}{15\, v}\left( 8\, m_e + 7 \frac{m_\mu + m_\tau}{2} \right) \simeq 0.00253. \nonumber
\end{align}
As can be seen from Eqs.~\eqref{eq:ycondA5}, a certain level of fine-tuning of the three VEVs is required to obtain the correct hierarchy of charged-leptons.\footnote{The parameters $\epsilon_i^2$ include different couplings of the mediators to the flavons and lepton fields, apart from $v_i^2$ and therefore are not necessarily positive.} Given that we have one undetermined parameter, we require no fine-tuning at least for the heaviest of the charged leptons, which implies that both $\epsilon_\xi$ and $3 \epsilon_{3'}^2$ should be of the same order; we choose 
 \begin{align}
    \epsilon_\xi\simeq 0.00126\,, \qquad \epsilon_{3'}^2\simeq 0.00042\,.
\end{align}

\subsection{``Shaping'' Symmetry}
As shown in Ref.~\cite{DiIura:2018fnk}, neutrino mixings are correctly reproduced if the neutrino flavons are invariant under a remaining ${\mc Z}_2 \times {\rm CP}$ symmetry. However, in this section we are only interested in the effect of neutrino flavons on the charged-lepton Yukawas. Following Ref. \cite{DiIura:2018fnk}, we assign the three right-handed neutrinos $\nu_R \equiv (\nu_{R_1}, \nu_{R_2}, \nu_{R_3})$ to a $\g{3}^\prime$ representation, which yields the following neutrino Lagrangian 
\begin{align}\label{eqn:CNuSuperPot}
	 {\cal L}^\nu \;=\; {\bar L}\nu_R\, H\, \frac{1}{\Lambda_f}\,\left( \varphi^\nu_\g{4} + \varphi^\nu_\g{5} \right) + \frac{1}{2}\, \nu_R\,\nu_R\, \frac{1}{\Lambda_f}\,\xi_\g{1},
\end{align}
when the corresponding $\mc Z_{n}$ charges are 
\begin{align} \label{ZnNeutrino}
    [{\bar L} \nu_R H] = p,\quad [\varphi^\nu_{\bf 4}] = [\varphi^\nu_{\bf 5}] = (n-p).
\end{align}
The neutrino mass matrix obtained from these Dirac and Majorana matrices is diagonalized with a Golden Ratio mixing matrix \cite{Datta:2003qg, Kajiyama:2007gx, Everett:2008et}.

Recall that the fields in the charged lepton sector have $\mc Z_n$ charges given by Eq.~\eqref{ZnChargedLepton}. The combined $A_5 \times \mc Z_n$ symmetry has to forbid neutrino flavons from contributing to the charged-lepton Yukawas at lower orders. The charged-leptons Yukawas receive contributions from the $\g{1}$, $\g{3}$ and $\g{5}$ representations. Singlets are not problematic, as they do not change the diagonal structure of the matrix. However, the off-diagonal Yukawa structure must be protected against bounds from lepton flavor violation. Neutrino mixings do not fix the size of neutrino-flavon VEVs, while, in a seesaw scenario, neutrino masses involve an additional scale, the right-handed neutrino mass; we are therefore free to choose the size of flavon VEVs with no effect on the standard neutrino parameters. We can assume the largest of the neutrino flavon VEVs to be ${\cal O }(\lambda^2)$. Then, we should forbid contributions from operators containing up to five flavon fields. For example, some of the lowest order contributions that must be prevented are $\varphi^\nu_\g{5}$ $(\varphi^\nu_\g{4},\varphi^\nu_\g{4})_\g{5}$,
$(\varphi^\nu_\g{4},\varphi^\nu_\g{5})_\g{3}$,
$(\varphi^\nu_\g{4},\varphi^\nu_\g{5})_\g{5}$,
$(\varphi^\nu_\g{5},\varphi^\nu_\g{5})_\g{5}$,
$((\varphi^\nu_\g{4},\varphi^\nu_\g{4}),\varphi^\nu_\g{4})_\g{5}$ etc, together with all possible combinations with daggered fields. 

An example of such a ``shaping'' symmetry is $\mc Z_{12}$, with the following charge assignment
\begin{equation}
\begin{gathered}
    {[\bar{L} \ell H] } = 6,\qquad [\xi] = [\varphi^e_{\g{3}}] = 6, \qquad [\varphi^e_{\g{5}}] = [\varphi^e_{\g{3'}}] = 3,\\
    [\bar{L} \nu_R H] = 7, \qquad [\varphi^\nu_{\g{4}}] = [\varphi^\nu_{\g{5}}] = 5.
\end{gathered}
\end{equation}
The charge assignment of the mediators can then be easily determined from the Feynman diagrams in Fig.~\ref{fig:A5Y}. We show the field content of the charged fermion sector and their charges in Table \ref{table:A5}.
\begin{table}[t]\centering
\renewcommand\arraystretch{1.1}
\begin{tabularx}{\textwidth}{@{}l  Y Y  Y  Y Y Y  Y Y Y Y Y Y @{}}
    \toprule
    Fields &  $\bar{L}$ & $\ell$ & $H$ & $\xi$ & $\varphi_{\g{3}}$ & $\varphi_{\g{3'}}$ & $\varphi_{\g{5}}$ & $\Omega$ & $\Sigma_{\g{3}}$ & $\Sigma_{\g{3'}}$ & $\Sigma_{\g{4}}$ & $\Sigma_{\g{5}}$  \\
    \midrule
    $SU(2)_L$ & $\g{2}$ & $\g{1}$ & $\g{2}$ & $\g{1}$ & $\g{1}$ & $\g{1}$ & $\g{1}$ & $\g{1}$ & $\g{1}$  & $\g{1}$ & $\g{1}$ & $\g{1}$  \\
    $A_5$ & $\g{3}_1$ & $\g{3}_1$ & $\g{1}$ & $\g{1}$ & $\g{3}$ & $\g{3'}$ & $\g{5}$ & $\g{3}$ & $\g{3}$  & $\g{3'}$ & $\g{4}$ & $\g{5}$  \\
    $\mc Z_{12}$ & $\g{\rho^{1}}$ &  $\g{\rho^{1}}$ &  $\g{\rho^{4}}$ &  $\g{\rho^{6}}$ &  $\g{\rho^{6}}$ &  $\g{\rho^{3}}$ &  $\g{\rho^{3}}$ &  $\g{\rho^{7}}$ &  $\g{\rho^{4}}$  & $\g{\rho^{4}}$ & $\g{\rho^{4}}$ &  $\g{\rho^{4}}$\\
    \bottomrule
\end{tabularx} 
\caption{Transformation properties of matter, scalar and messenger fields in the charged lepton sector. Here  $\rho^{12} = 1$. The $\mc Z_{12}$ `shaping' symmetry prevents the mixing of neutrino sector flavons with charged lepton sector fields and vice versa.
}
\label{table:A5}
\end{table}

\subsection{Predictions}
\begin{figure}[!h]
    \centering
    \includegraphics[scale=1]{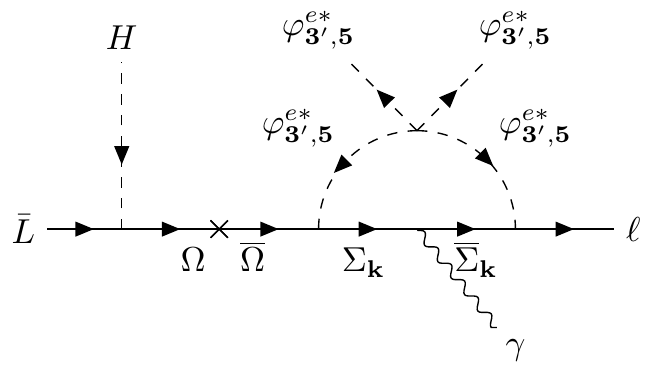}
    \caption{Diagram contributing to dipole operator.}
    \label{fig:A5dipole}
\end{figure}

The dipole matrix is generated by loop diagrams similar to Fig.~\ref{fig:loop}, except with an external photon line connected to the loop, as shown in Fig.~\ref{fig:A5dipole}. 
In contrast with the $\mc T_{13}$ model, a mixed quartic coupling in the scalar potential is not strictly necessary in order to obtain the correct sign of $(g-2)_\mu$, and we can limit ourselves to considering only self-couplings to close the loops. In fact, having several independent diagrams in Fig.~\ref{fig:A5Y} contributing to the different entries in the Yukawa matrix, we always have the freedom to require $Y_{\mu\mu}+\delta Y_{\mu\mu}>0$ but $\delta Y_{\mu\mu}<0$ at the price of a level of tuning.  
The dipole matrix is directly derived from the radiative mass contribution as $C_\ell =2 \pi^2 \delta Y_\ell (f_1\rightarrow f_2) /\Lambda_f^2 $:
\begin{align}
C_\ell = \frac{f_2(x_\varphi)}{16\, \Lambda_f^2}
\left(\begin{array}{ccc}
3\, \epsilon_{3'}^2\, \beta_{3'}\, y_3 + 7\, \epsilon_5^2\, \beta_5\, y_5 &  0  &  0 \\
0 & 3\, \epsilon_{3'}^2\, \beta_{3'}\, y_3 - 8\, \epsilon_5^2\, \beta_5\, y_5  &  0 \\
0 &  0  & 3\, \epsilon_{3'}^2\, \beta_{3'}\, y_3 - 8\, \epsilon_5^2\, \beta_5\, y_5 \\
\end{array}
\right), \label{eq:dYl3exact}
\end{align}
where $f_2(x_\varphi)$ is the loop function given by Eq.~\eqref{eq:f2}.
The structure has $C_{\tau\tau}=C_{\mu\mu}$ which automatically predicts the anomalous magnetic moment of the $\tau$ lepton to be 
\begin{align}
\Delta a_\tau=\frac{m_\tau}{m_\mu}\Delta a_\mu=4.24~\times~10^{-8}.
\end{align}
Requiring $C_{\mu\mu}$ to be in the range compatible with the observed discrepancy in Table~\ref{table:DipoleConst},
\begin{equation}
    M_\chi \sim \Lambda_f = \frac{1}{4\pi}\sqrt{\frac{15\, \epsilon_5^2\, \beta_{5}\, v\, m_e\, m_\mu }{2\left(\,m_e\Delta a_\mu + m_\mu \Delta a_e\right)}f_2(x_{\varphi_5})}\,.
\end{equation}
If we consider $\beta_{5}=2 \pi$ and $\epsilon^2_{5}$ as in Eq.~(\ref{eq:ycondA5}) we can derive the maximum value for the mediator mass scale allowed by the model, which reads $\Lambda_f\leq 844$ GeV at the 1$\sigma$ level and  $\Lambda_f\leq 1354$ GeV at 3$\sigma$. Higher values of $\Lambda_f$ can be reached by also allowing for a level of fine-tuning in the $\tau$-mass.  

\section{Comments and Outlook} \label{sec:outlook}
We have seen in the previous sections that it is still possible to explain the observed discrepancy in the muon anomalous magnetic moment through a low-scale flavor symmetry in the leptonic sector. Nevertheless, given that the dipole matrix for charged leptons is not expected to be exactly proportional to the Yukawa matrix in these models, the
strong constraints from LFV processes require the charged-lepton Yukawa couplings to be (quasi-)diagonal in the flavor basis, up to a very high order in the flavon fields. This implies that the different operators for the $e$, $\mu$ and $\tau$ masses up to, at least, order $\lambda^8$, contribute only to diagonal Yukawa elements in the same basis after replacing the flavon VEVs.

This is not simple and, in fact, is not possible for most flavor models in the literature. The observation of non-vanishing neutrino mixings implies that the charged-lepton and neutrino mass matrices can not be made simultaneously diagonal in a single basis. When the mass matrices are expressed in terms of flavon VEVs (or products of VEVs), this implies that some flavons must contribute to the off-diagonal elements. If all flavons couple to both the charged-lepton and neutrino sector, both matrices will always have off-diagonal entries. 

This is indeed what happens in most models aiming to simultaneously describe the quark and lepton Yukawa matrices, perhaps pointing to a grand unification scenario. In these scenarios, up- and down-quark Yukawa matrices are off-diagonal, as well as the charged-lepton and neutrino Yukawas (see, for instance \cite{King:2001uz,King:2003rf,Ross:2004qn,deMedeirosVarzielas:2005ax,Calibbi:2008qt,deMedeirosVarzielas:2017sdv,DeMedeirosVarzielas:2018dyz}). Taking as an example the $\Delta (27)$ model of Ref.~\cite{DeMedeirosVarzielas:2018dyz}, we have only three different flavons that break $\Delta(27)$ and provide the structure of the mass matrices with the following alignment
\begin{align}
\langle \varphi_{c} \rangle \propto  \left(\begin{array}{c} 0 \\ 0\\  1\end{array} \right) \sim \langle \varphi\rangle \hspace{3mm},\hspace{3mm}\langle \varphi_{b} \rangle \propto \cfrac{1}{\sqrt{2}}\, \left(\begin{array}{c} 0 \\ 1\\  1\end{array} \right)\,,\hspace{5mm}\langle \varphi_{a} \rangle\propto \cfrac{1}{\sqrt{3}}\, \left(\begin{array}{c} 1 \\ 1\\  -1\end{array} \right). 
\end{align}
The core prediction of the model is complex symmetric Dirac and Majorana mass matrices with a universal texture zero (UTZ) in the $(11)$ entry for all fermion families, which at the lowest order are of the form
\\
\begin{align}
\label{eq:Mfeps}
Y_{e,\nu}\,\propto\,y_c^{e,\nu}
\begin{pmatrix}
0 &  \,\epsilon_{e,\nu}^3 & \,\epsilon_{e,\nu}^3 \\[6pt]
\,\epsilon_{e,\nu}^3 &  \,\epsilon_{e,\nu}^2 &  \,\epsilon_{e,\nu}^2\\[6pt]
\,\epsilon_{e,\nu}^3 &  \,\epsilon_{e,\nu}^2 & 1\\ 
\end{pmatrix}  \,,
\end{align}
with $\epsilon_e \simeq 0.15$ and 
$\epsilon_\nu \simeq 0.05$.

Clearly, if we take into account radiative corrections involving flavons, this model would not be able to explain the observed discrepancy of $(g-2)_\mu$ that requires a mediator mass below one TeV. The constraints from LFV processes would push the mass of the mediator to a scale $\Lambda>14$ TeV and therefore the contribution to  $(g-2)_\mu$ would be $\Delta a_\mu \lesssim 10^{-13}$.

A partial solution to this problem is the ``sequestering" mechanism that we have used in the previous sections. In this mechanism the flavor symmetry, ${\cal G}_f$, is completely broken by the full set of matrices $M_e$ and $M_\nu$, but $M_e$ and $M_\nu$ are separately invariant under some nontrivial subgroups ${\cal G}_e$ and ${\cal G}_\nu$. This is done by dividing the flavons in two different subsets, $\varphi_e$ and $\varphi_\nu$, breaking ${\cal G}_f$ to ${\cal G}_e$ and ${\cal G}_\nu$, respectively, and coupling only, at leading order, to charged-leptons or neutrinos. However, it is not possible to completely forbid $\varphi_e$ and $\varphi_\nu$ from coupling to both sectors. This follows as the sequestering is obtained through a global discrete $\mathcal{Z}_n$ or even a continuous abelian symmetry; for a high enough order in the numbers of flavons we can always obtain zero or $n$ total charge, and, hence, both fields will be able to couple to the opposite sector. 

To be precise, for flavor symmetries reproducing the observed neutrino mixings, it is enough to protect sequestering up to the size of the experimental errors on the mixings. \textcolor{black}{This means that, if we accept an error on neutrino mixings of order $\delta\sin \theta_{12} /\sin \theta_{12}  \lesssim 0.03 \sim \lambda^2 $, the corrections to the charged-lepton Yukawa $Y_{\ell}^{12}$ can be of the order $\lambda^4 Y_{\ell}^{33}$, a factor $\lambda^2$ smaller than the diagonal element $Y_\ell^{22}$.}  Again, this would not satisfy the bounds of Eq.~\eqref{eq:Ye-bounds}, and this stronger bound would override the bound from the anomalous magnetic moment on the flavor scale. The precise bound on the flavor scale will depend on the mechanism protecting the sequestering in each model. Unfortunately,  this mechanism is not specified in many of the models that consider only leading order predictions. In general, if we assume that the corrections to the charged lepton Yukawas are those allowed by the experimental errors on $\sin{\theta_{12}}$, the bound on the flavor scale would be $\Lambda_f > 6.4$ TeV and the contribution to the muon $g-2$ is $\Delta a_{\mu}\lesssim 4.9 \times 10^{-13} $.

Therefore, we can see that lepton flavor violating processes place very strong constraints on the structure of the dipole matrix, which translates into very strong constraints on the scale of flavor symmetry breaking and mediator masses. Only models which protect the diagonal structure of the charged-lepton Yukawas in the basis of flavor interactions up to a very high-orders are able to explain the muon anomalous magnetic moment consistent with LFV bounds. In most models, it is not possible to completely reproduce the observed discrepancy and the contribution from the flavor symmetry is typically negligible.

\section{Conclusion} \label{sec:conclusions}
The recent observation of the $4.2\sigma$ discrepancy in the muon anomalous magnetic moment at Fermilab strengthens our expectation for observing new physics at work around the TeV scale. In this paper we have pursued the idea that this new physics contribution could come entirely from the breaking of a low-scale flavor symmetry, and have built two explicit models based on the flavor groups $\mc T_{13}$ and $A_5$ to demonstrate this.

The Yukawa matrix describing charged lepton masses and mixings has the same flavor structure as the dipole matrix yielding lepton anomalous magnetic moments and flavor violation. However, because of $\mc O(1)$ differences in the coefficients of their entries in the flavor basis, the dipole matrix is not diagonalized along with the Yukawa. The off-diagonal entries of the dipole matrix in the mass basis are, up to a global loop factor, at least of the same order as the corresponding entries in the flavor basis, barring accidental cancellations. Stringent constraints from non-observation of lepton flavor violation strictly restricts these entries, and in turn, the corresponding off-diagonal entries of the Yukawa matrix in the flavor basis. Effectively, the charged lepton Yukawas need to be (quasi-)diagonal up to very high orders in the Wolfenstein parameter $\lambda$ to evade LFV constraints. 

It is very challenging to build models based on flavor symmetries which can account for the entirety of the discrepancy in the muon $(g-2)$ while being consistent with LFV restrictions. We have demonstrated that the flavor groups $\mc T_{13}$ and $A_5$ are suitable in this regard, and have built two models that fulfill our objectives when the flavor symmetries are broken below the TeV scale. 

For the $\mc T_{13}$ model, the Yukawa matrix in the flavor basis is quasi-diagonal, where the vanishing $(11)$ entry is compensated by the nonzero $(13)-(31)$ entries. This gives a nonzero contribution to $\tau \rightarrow e\gamma$ which is within experimental bounds and within reach of near future experiments \cite{Aushev:2010bq}. This model accounts for the observed discrepancy in $(g-2)_\mu$ at $1\sigma$ ($3\sigma$) level when the flavor symmetry is broken at $430$ GeV ($705$ GeV). 

The $A_5$ model yields a diagonal Yukawa structure in the limit of exact sequestering. This diagonal structure can be protected with the help of an appropriate discrete symmetry. In this way, it is possible to prohibit all LFV processes to the required level and reproduce the contribution to $(g-2)_\mu$ at the $1\sigma$ ($3\sigma$) level when the symmetry breaking scale is $844$ GeV ($1354$ GeV).

In both cases, the effect of symmetry breaking is transferred to the Standard Model fermions via vectorlike mediators, which could transform as doublets or singlets under the Standard Model $SU(2)$, depending on the construction of the Feynman diagrams. We have chosen these mediators to be $SU(2)$ singlets to avoid tight lower bounds on doublet vectorlike leptons from collider searches. 

The requirement of having a (quasi-)diagonal Yukawa structure to elude LFV constraints  and simultaneously emulating the observed discrepancy in $(g-2)_\mu$ is very challenging, and very few flavor symmetries have the requisite structure that can facilitate this without fine-tuning. We have identified two such symmetries and have discussed briefly why many of the popular flavor symmetries in literature would fail to achieve this feat. It would be interesting to expand the models discussed in this paper to the quark and the neutrino sector, which we leave for a future study.

\section*{Acknowledgments}
We thank Lorenzo Calibbi for useful discussion in the initial stage of this work. M.L.L.I, A.M., M.J.P and O.V. acknowledge support from Spanish AEI-MICINN, PID2020-113334GB-I00/AEI/10.13039/501100011033. 
M.L.L.I. acknowledges partial support from Plan Propio de Investigación 2021 de la Universidad de Córdoba.
M.H.R. acknowledges partial support from  U.S. Department of Energy under Grant No. DE-SC0010296. O.V. was partially supported by ``Generalitat Valenciana'' grant PROMETEO2017-033. A.M. was supported by the Estonian Research Council grant MOBTT86 “Probing the Higgs sector at the LHC and beyond”.
\appendix

\section{$\mc T_{13}$ Clebsch-Gordan Coefficients} \label{app:T13}
In this appendix, we list the Clebsch-Gordan coefficients of $\mc T_{13}$ following Ref.~\cite{ishimori2012introduction}.
\begin{align*}
\matc{
\ket{1}\\
\ket{2}\\
\ket{3}
}_{\g{3}_1}
\otimes
\matc{
\ket{1'}\\
\ket{2'}\\
\ket{3'}
}_{\g{3}_1} &= 
\matc{
\ket{1} \ket{1'}\\
\ket{2} \ket{2'}\\
\ket{3} \ket{3'}
}_{\g{3}_2} \oplus
\matc{
\ket{2} \ket{3'}\\
\ket{3} \ket{1'}\\
\ket{1} \ket{2'}
}_{\gb{3}_1} \oplus
\matc{
\ket{3} \ket{2'}\\
\ket{1} \ket{3'}\\
\ket{2} \ket{1'}
}_{\gb{3}_1} \\
\matc{
\ket{1}\\
\ket{2}\\
\ket{3}
}_{\g{3}_2}
\otimes
\matc{
\ket{1'}\\
\ket{2'}\\
\ket{3'}
}_{\g{3}_2} &= 
\matc{
\ket{2} \ket{2'}\\
\ket{3} \ket{3'}\\
\ket{1} \ket{1'}
}_{\gb{3}_1} \oplus
\matc{
\ket{2} \ket{3'}\\
\ket{3} \ket{1'}\\
\ket{1} \ket{2'}
}_{\gb{3}_2} \oplus
\matc{
\ket{3} \ket{2'}\\
\ket{1} \ket{3'}\\
\ket{2} \ket{1'}
}_{\gb{3}_2}\\
\matc{
\ket{1}\\
\ket{2}\\
\ket{3}
}_{\g{3}_1}
\otimes
\matc{
\ket{1'}\\
\ket{2'}\\
\ket{3'}
}_{\g{3}_2} &= 
\matc{
\ket{3} \ket{3'}\\
\ket{1} \ket{1'}\\
\ket{2} \ket{2'}
}_{\g{3}_1} \oplus
\matc{
\ket{3} \ket{1'}\\
\ket{1} \ket{2'}\\
\ket{2} \ket{3'}
}_{\gb{3}_2} \oplus
\matc{
\ket{3} \ket{2'}\\
\ket{1} \ket{3'}\\
\ket{2} \ket{1'}
}_{\g{3}_2}
\\
\matc{
\ket{1}\\
\ket{2}\\
\ket{3}
}_{\g{3}_1}
\otimes
\matc{
\ket{1'}\\
\ket{2'}\\
\ket{3'}
}_{\gb{3}_2} &= 
\matc{
\ket{1} \ket{1'}\\
\ket{2} \ket{2'}\\
\ket{3} \ket{3'}
}_{\gb{3}_1} \oplus
\matc{
\ket{2} \ket{3'}\\
\ket{3} \ket{1'}\\
\ket{1} \ket{2'}
}_{\gb{3}_2} \oplus
\matc{
\ket{2} \ket{1'}\\
\ket{3} \ket{2'}\\
\ket{1} \ket{3'}
}_{\g{3}_1}\\
\matc{
\ket{1}\\
\ket{2}\\
\ket{3}
}_{\g{3}_2}
\otimes
\matc{
\ket{1'}\\
\ket{2'}\\
\ket{3'}
}_{\gb{3}_1} &= 
\matc{
\ket{1} \ket{1'}\\
\ket{2} \ket{2'}\\
\ket{3} \ket{3'}
}_{\g{3}_1} \oplus
\matc{
\ket{1} \ket{2'}\\
\ket{2} \ket{3'}\\
\ket{3} \ket{1'}
}_{\gb{3}_1} \oplus
\matc{
\ket{3} \ket{2'}\\
\ket{1} \ket{3'}\\
\ket{2} \ket{1'}
}_{\g{3}_2}
\\
\matc{
\ket{1}\\
\ket{2}\\
\ket{3}
}_{\g{3}_1}
\otimes
\matc{
\ket{1'}\\
\ket{2'}\\
\ket{3'}
}_{\gb{3}_1} &= 
\matc{
\ket{1} \ket{2'}\\
\ket{2} \ket{3'}\\
\ket{3} \ket{1'}
}_{\gb{3}_2} \oplus
\matc{
\ket{2} \ket{1'}\\
\ket{3} \ket{2'}\\
\ket{1} \ket{3'}
}_{\g{3}_2} \\
&\oplus  (\ket{1}\ket{1'}+\ket{2}\ket{2'}+\ket{3}\ket{3'})_\g{1}\\
&\oplus  (\ket{1}\ket{1'}+\omega \ket{2}\ket{2'}+ \omega^2 \ket{3}\ket{3'})_\g{1'}\\
&\oplus (\ket{1}\ket{1'}+\omega^2 \ket{2}\ket{2'}+ \omega \ket{3}\ket{3'})_{\gb{1}'}
\\
\matc{
\ket{1}\\
\ket{2}\\
\ket{3}
}_{\g{3}_2}
\otimes
\matc{
\ket{1'}\\
\ket{2'}\\
\ket{3'}
}_{\gb{3}_2} &= 
\matc{
\ket{2} \ket{3'}\\
\ket{3} \ket{1'}\\
\ket{1} \ket{2'}
}_{\g{3}_1} \oplus
\matc{
\ket{3} \ket{2'}\\
\ket{1} \ket{3'}\\
\ket{2} \ket{1'}
}_{\gb{3}_1} \\
&\oplus  (\ket{1}\ket{1'}+\ket{2}\ket{2'}+\ket{3}\ket{3'})_\g{1}\\
&\oplus  (\ket{1}\ket{1'}+\omega \ket{2}\ket{2'}+ \omega^2 \ket{3}\ket{3'})_\g{1'}\\
&\oplus (\ket{1}\ket{1'}+\omega^2 \ket{2}\ket{2'}+ \omega \ket{3}\ket{3'})_{\gb{1}'}, \qquad \omega^3 = 1
\end{align*}

\newpage

\section{$A_5$ Clebsch-Gordan Coefficients} \label{app:A5}
In this appendix we list the $A_5$ Clebsch-Gordan coefficients following Ref.~\cite{Ding:2011cm}.
\begingroup
\allowdisplaybreaks
\begin{align*}
\matc{
\ket{1}\\
\ket{2}\\
\ket{3}
}_{\g{3}}
\otimes
\matc{
\ket{1'}\\
\ket{2'}\\
\ket{3'}
}_{\g{3}} &= \left(|1\rangle |1'\rangle+|2\rangle |3'\rangle+|3\rangle |2'\rangle\right)_{\g{1}_S}\\ &\oplus\left(\begin{array}{c}
|2\rangle |3'\rangle-|3\rangle |2'\rangle \\
|1\rangle |2'\rangle-|2\rangle |1'\rangle \\
|3\rangle |1'\rangle-|1\rangle |3'\rangle
\end{array}\right)_{\mathbf{3}_{A}} 
\oplus \left(\begin{array}{c}
2 |1\rangle |1'\rangle-|2\rangle |3'\rangle-|3\rangle |2'\rangle \\
-\sqrt{3} |1\rangle |2'\rangle-\sqrt{3} |2\rangle |1'\rangle \\
\sqrt{6} |2\rangle |2'\rangle \\
\sqrt{6} |3\rangle |3'\rangle \\
-\sqrt{3} |1\rangle |3'\rangle-\sqrt{3} |3\rangle |1'\rangle
\end{array}\right)_{\mathbf{5}_{S}}
\\
\matc{
\ket{1}\\
\ket{2}\\
\ket{3}
}_{\g{3'}}
\otimes
\matc{
\ket{1'}\\
\ket{2'}\\
\ket{3'}
}_{\g{3'}} &=  \left(|1\rangle |1'\rangle+|2\rangle |3'\rangle+|3\rangle |2'\rangle\right)_{\g{1}_S} \\
&\oplus\left(\begin{array}{c}
|2\rangle |3'\rangle-|3\rangle |2'\rangle \\
|1\rangle |2'\rangle-|2\rangle |1'\rangle \\
|3\rangle |1'\rangle-|1\rangle |3'\rangle
\end{array}\right)_{\mathbf{3'}_{A}} 
\oplus \left(\begin{array}{c}
2 \ket{1} \ket{1'}-\ket{2} \ket{3'}-\ket{3} \ket{2'} \\
\sqrt{6} \ket{3} \ket{3'} \\
-\sqrt{3} \ket{1} \ket{2'}-\sqrt{3} \ket{2} \ket{1'} \\
-\sqrt{3} \ket{1} \ket{3'}-\sqrt{3} \ket{3} \ket{1'} \\
\sqrt{6} \ket{2} \ket{2'}
\end{array}\right)_{\mathbf{5}_{S}} 
\\
\matc{
\ket{1}\\
\ket{2}\\
\ket{3}
}_{\g{3}}
\otimes
\matc{
\ket{1'}\\
\ket{2'}\\
\ket{3'}
}_{\g{3'}} &= \left(\begin{array}{c}
\sqrt{2} \ket{2} \ket{1'}+\ket{3} \ket{2'} \\
-\sqrt{2} \ket{1} \ket{2'}-\ket{3} \ket{3'} \\
-\sqrt{2} \ket{1} \ket{3'}-\ket{2} \ket{2'} \\
\sqrt{2} \ket{3} \ket{1'}+\ket{2} \ket{3'}
\end{array}\right)_{\g{4}} \oplus \left(\begin{array}{c}
\sqrt{3} \ket{1} \ket{1'} \\
\ket{2} \ket{1'}-\sqrt{2} \ket{3} \ket{2'} \\
\ket{1} \ket{2'}-\sqrt{2} \ket{3} \ket{3'} \\
\ket{1} \ket{3'}-\sqrt{2} \ket{2} \ket{2'} \\
\ket{3} \ket{1'}-\sqrt{2} \ket{2} \ket{3'}
\end{array}\right)_{\g{5}} 
\\
\matc{
\ket{1}\\
\ket{2}\\
\ket{3}
}_{\g{3}}
\otimes
\matc{
\ket{1'}\\
\ket{2'}\\
\ket{3'}\\
\ket{4'}
}_{\g{4}} &= \left(\begin{array}{c}
-\sqrt{2} \ket{2} \ket{4'}-\sqrt{2} \ket{3} \ket{1'} \\
\sqrt{2} \ket{1} \ket{2'}-\ket{2} \ket{1'}+\ket{3} \ket{3'} \\
\sqrt{2} \ket{1} \ket{3'}+\ket{2} \ket{2'}-\ket{3} \ket{4'}
\end{array}\right)_{\g{3'}} \oplus \left(\begin{array}{c}
\ket{1} \ket{1'}-\sqrt{2} \ket{3} \ket{2'} \\
-\ket{1} \ket{2'}-\sqrt{2} \ket{2} \ket{1'} \\
\ket{1} \ket{3'}+\sqrt{2} \ket{3} \ket{4'} \\
-\ket{1} \ket{4'}+\sqrt{2} \ket{2} \ket{3'}
\end{array}\right)_{\g{4}} \\
&\oplus \left(\begin{array}{c}
\sqrt{6} \ket{2} \ket{4'}-\sqrt{6} \ket{3} \ket{1'} \\
2 \sqrt{2} \ket{1} \ket{1'}+2 \ket{3} \ket{2'} \\
-\sqrt{2} \ket{1} \ket{2'}+\ket{2} \ket{1'}+3 \ket{3} \ket{3'} \\
\sqrt{2} \ket{1} \ket{3'}-3 \ket{2} \ket{2'}-\ket{3} \ket{4'} \\
-2 \sqrt{2} \ket{1} \ket{4'}-2 \ket{2} \ket{3'}
\end{array}\right)_{\g{5}}\\
\matc{
\ket{1}\\
\ket{2}\\
\ket{3}
}_{\g{3'}}
\otimes
\matc{
\ket{1'}\\
\ket{2'}\\
\ket{3'}\\
\ket{4'}
}_{\g{4}} &= \left(\begin{array}{c}
-\sqrt{2} \ket{2} \ket{3'}-\sqrt{2} \ket{3} \ket{2'} \\
\sqrt{2} \ket{1} \ket{1'}+\ket{2} \ket{4'}-\ket{3} \ket{3'} \\
\sqrt{2} \ket{1} \ket{4'}-\ket{2} \ket{2'}+\ket{3} \ket{1'}
\end{array}\right)_{\g{3}} \oplus \left(\begin{array}{c}
\ket{1} \ket{1'}+\sqrt{2} \ket{3} \ket{3'} \\
\ket{1} \ket{2'}-\sqrt{2} \ket{3} \ket{4'} \\
-\ket{1} \ket{3'}+\sqrt{2} \ket{2} \ket{1'} \\
-\ket{1} \ket{4'}-\sqrt{2} \ket{2} \ket{2'}
\end{array}\right)_{\g{4}} \\
&\oplus \left(\begin{array}{c}
\sqrt{6} \ket{2} \ket{3'}-\sqrt{6} \ket{3} \ket{2'} \\
\sqrt{2} \ket{1} \ket{1'}-3 \ket{2} \ket{4'}-\ket{3} \ket{3'} \\
2 \sqrt{2} \ket{1} \ket{2'}+2 \ket{3} \ket{4'} \\
-2 \sqrt{2} \ket{1} \ket{3'}-2 \ket{2} \ket{1'} \\
-\sqrt{2} \ket{1} \ket{4'}+\ket{2} \ket{2'}+3 \ket{3} \ket{1'}
\end{array}\right)_{\g{5}}
\\
\matc{
\ket{1}\\
\ket{2}\\
\ket{3}
}_{\g{3}}
\otimes
\matc{
\ket{1'}\\
\ket{2'}\\
\ket{3'}\\
\ket{4'}\\
\ket{5'}
}_{\g{5}} &= \left(\begin{array}{c}
-2 \ket{1} \ket{1'}+\sqrt{3} \ket{2} \ket{5'}+\sqrt{3} \ket{3} \ket{2'} \\
\sqrt{3} \ket{1} \ket{2'}+\ket{2} \ket{1'}-\sqrt{6} \ket{3} \ket{3'} \\
\sqrt{3} \ket{1} \ket{5'}-\sqrt{6} \ket{2} \ket{4'}+\ket{3} \ket{1'}
\end{array}\right)_{\g{3}}\\
&\oplus \left(\begin{array}{c}
\sqrt{3} \ket{1} \ket{1'}+\ket{2} \ket{5'}+\ket{3} \ket{2'} \\
\ket{1} \ket{3'}-\sqrt{2} \ket{2} \ket{2'}-\sqrt{2} \ket{3} \ket{4'} \\
\ket{1} \ket{4'}-\sqrt{2} \ket{2} \ket{3'}-\sqrt{2} \ket{3} \ket{5'}
\end{array}\right)_{\g{3'}} \\
&\oplus \left(\begin{array}{c}
2 \sqrt{2} \ket{1} \ket{2'}-\sqrt{6} \ket{2} \ket{1'}+\ket{3} \ket{3'} \\
-\sqrt{2} \ket{1} \ket{3'}+2 \ket{2} \ket{2'}-3 \ket{3} \ket{4'} \\
\sqrt{2} \ket{1} \ket{4'}+3 \ket{2} \ket{3'}-2 \ket{3} \ket{5'} \\
-2 \sqrt{2} \ket{1} \ket{5'}-\ket{2} \ket{4'}+\sqrt{6} \ket{3} \ket{1'}
\end{array}\right)_{\g{4}}\\
&\oplus \left(\begin{array}{c}
\sqrt{3} \ket{2} \ket{5'}-\sqrt{3} \ket{3} \ket{2'} \\
-\ket{1} \ket{2'}-\sqrt{3} \ket{2} \ket{1'}-\sqrt{2} \ket{3} \ket{3'} \\
-2 \ket{1} \ket{3'}-\sqrt{2} \ket{2} \ket{2'} \\
2 \ket{1} \ket{4'}+\sqrt{2} \ket{3} \ket{5'} \\
\ket{1} \ket{5'}+\sqrt{2} \ket{2} \ket{4'}+\sqrt{3} \ket{3} \ket{1'}
\end{array}\right)_{\g{5}}\\
\matc{
\ket{1}\\
\ket{2}\\
\ket{3}
}_{\g{3'}}
\otimes
\matc{
\ket{1'}\\
\ket{2'}\\
\ket{3'}\\
\ket{4'}\\
\ket{5'}
}_{\g{5}} &= \left(\begin{array}{c}
\sqrt{3} \ket{1} \ket{1'}+\ket{2} \ket{4'}+\ket{3} \ket{3'} \\
\ket{1} \ket{2'}-\sqrt{2} \ket{2} \ket{5'}-\sqrt{2} \ket{3} \ket{4'} \\
\ket{1} \ket{5'}-\sqrt{2} \ket{2} \ket{3'}-\sqrt{2} \ket{3} \ket{2'}
\end{array}\right)_{\g{3}}\\
&\oplus \left(\begin{array}{c}
-2 \ket{1} \ket{1'}+\sqrt{3} \ket{2} \ket{4'}+\sqrt{3} \ket{3} \ket{3'} \\
\sqrt{3} \ket{1} \ket{3'}+\ket{2} \ket{1'}-\sqrt{6} \ket{3} \ket{5'} \\
\sqrt{3} \ket{1} \ket{4'}-\sqrt{6} \ket{2} \ket{2'}+\ket{3} \ket{1'}
\end{array}\right)_{\g{3'}} \\
&\oplus \left(\begin{array}{c}
\sqrt{2} \ket{1} \ket{2'}+3 \ket{2} \ket{5'}-2 \ket{3} \ket{4'} \\
2 \sqrt{2} \ket{1} \ket{3'}-\sqrt{6} \ket{2} \ket{1'}+\ket{3} \ket{5'} \\
-2 \sqrt{2} \ket{1} \ket{4'}-\ket{2} \ket{2'}+\sqrt{6} \ket{3} \ket{1'} \\
-\sqrt{2} \ket{1} \ket{5'}+2 \ket{2} \ket{3'}-3 \ket{3} \ket{2'}
\end{array}\right)_{\g{4}}\\
&\oplus \left(\begin{array}{c}
\sqrt{3} \ket{2} \ket{4'}-\sqrt{3} \ket{3} \ket{3'} \\
2 \ket{1} \ket{2'}+\sqrt{2} \ket{3} \ket{4'} \\
-\ket{1} \ket{3'}-\sqrt{3} \ket{2} \ket{1'}-\sqrt{2} \ket{3} \ket{5'} \\
\ket{1} \ket{4'}+\sqrt{2} \ket{2} \ket{2'}+\sqrt{3} \ket{3} \ket{1'} \\
-2 \ket{1} \ket{5'}-\sqrt{2} \ket{2} \ket{3'}
\end{array}\right)_{\g{5}}
\\
\matc{
\ket{1}\\
\ket{2}\\
\ket{3}\\
\ket{4}
}_{\g{4}}
\otimes
\matc{
\ket{1'}\\
\ket{2'}\\
\ket{3'}\\
\ket{4'}
}_{\g{4}} &=  (\ket{1} \ket{4'}+\ket{2} \ket{3'}+\ket{3} \ket{2'}+\ket{4} \ket{1'})_{\g{1}_S} \\
&\oplus\left(\begin{array}{c}
-\ket{1} \ket{4'}+\ket{2} \ket{3'}-\ket{3} \ket{2'}+\ket{4} \ket{1'} \\
\sqrt{2} \ket{2} \ket{4'}-\sqrt{2} \ket{4} \ket{2'} \\
\sqrt{2} \ket{1} \ket{3'}-\sqrt{2} \ket{3} \ket{1'}
\end{array}\right)_{\g{3}_A} \\
&\oplus \left(\begin{array}{c}
\ket{1} \ket{4'}+\ket{2} \ket{3'}-\ket{3} \ket{2'}-\ket{4} \ket{1'} \\
\sqrt{2} \ket{3} \ket{4'}-\sqrt{2} \ket{4} \ket{3'} \\
\sqrt{2} \ket{1} \ket{2'}-\sqrt{2} \ket{2} \ket{1'}
\end{array}\right)_{\g{3'}_A}\\
&\oplus \left(\begin{array}{c}
\ket{2} \ket{4'}+\ket{3} \ket{3'}+\ket{4} \ket{2'} \\
\ket{1} \ket{1'}+\ket{3} \ket{4'}+\ket{4} \ket{3'} \\
\ket{1} \ket{2'}+\ket{2} \ket{1'}+\ket{4} \ket{4'} \\
\ket{1} \ket{3'}+\ket{2} \ket{2'}+\ket{3} \ket{1'}
\end{array}\right)_{\g{4}_S} \\
&\oplus \left(\begin{array}{c}
\sqrt{3} \ket{1} \ket{4'}-\sqrt{3} \ket{2} \ket{3'}-\sqrt{3} \ket{3} \ket{2'}+\sqrt{3} \ket{4} \ket{1'} \\
-\sqrt{2} \ket{2} \ket{4'}+2 \sqrt{2} \ket{3} \ket{3'}-\sqrt{2} \ket{4} \ket{2'} \\
-2 \sqrt{2} \ket{1} \ket{1'}+\sqrt{2} \ket{3} \ket{4'}+\sqrt{2} \ket{4} \ket{3'} \\
\sqrt{2} \ket{1} \ket{2'}+\sqrt{2} \ket{2} \ket{1'}-2 \sqrt{2} \ket{4} \ket{4'} \\
-\sqrt{2} \ket{1} \ket{3'}+2 \sqrt{2} \ket{2} \ket{2'}-\sqrt{2} \ket{3} \ket{1'}
\end{array}\right)_{\g{5}_S}
\\
\matc{
\ket{1}\\
\ket{2}\\
\ket{3}\\
\ket{4}
}_{\g{4}}
\otimes
\matc{
\ket{1'}\\
\ket{2'}\\
\ket{3'}\\
\ket{4'}\\
\ket{5'}
}_{\g{5}} &=   \left(\begin{array}{c}
2 \sqrt{2} \ket{1} \ket{5'}-\sqrt{2} \ket{2} \ket{4'}+\sqrt{2} \ket{3} \ket{3'}-2 \sqrt{2} \ket{4} \ket{2'} \\
-\sqrt{6} \ket{1} \ket{1'}+2 \ket{2} \ket{5'}+3 \ket{3} \ket{4'}-\ket{4} \ket{3'} \\
\ket{1} \ket{4'}-3 \ket{2} \ket{3'}-2 \ket{3} \ket{2'}+\sqrt{6} \ket{4} \ket{1'}
\end{array}\right)_{\g{3}} \\
&\oplus \left(\begin{array}{c}
\sqrt{2} \ket{1} \ket{5'}+2 \sqrt{2} \ket{2} \ket{4'}-2 \sqrt{2} \ket{3} \ket{3'}-\sqrt{2} \ket{4} \ket{2'} \\
3 \ket{1} \ket{2'}-\sqrt{6} \ket{2} \ket{1'}-\ket{3} \ket{5'}+2 \ket{4} \ket{4'} \\
-2 \ket{1} \ket{3'}+\ket{2} \ket{2'}+\sqrt{6} \ket{3} \ket{1'}-3 \ket{4} \ket{5'}
\end{array}\right)_{\g{3'}} \\
&\oplus \left(\begin{array}{c}
\sqrt{3} \ket{1} \ket{1'}-\sqrt{2} \ket{2} \ket{5'}+\sqrt{2} \ket{3} \ket{4'}-2 \sqrt{2} \ket{4} \ket{3'} \\
-\sqrt{2} \ket{1} \ket{2'}-\sqrt{3} \ket{2} \ket{1'}+2 \sqrt{2} \ket{3} \ket{5'}+\sqrt{2} \ket{4} \ket{4'} \\
\sqrt{2} \ket{1} \ket{3'}+2 \sqrt{2} \ket{2} \ket{2'}-\sqrt{3} \ket{3} \ket{1'}-\sqrt{2} \ket{4} \ket{5'} \\
-2 \sqrt{2} \ket{1} \ket{4'}+\sqrt{2} \ket{2} \ket{3'}-\sqrt{2} \ket{3} \ket{2'}+\sqrt{3} \ket{4} \ket{1'}
\end{array}\right)_{\g{4}} \\
&\oplus \left(\begin{array}{c}
\sqrt{2} \ket{1} \ket{5'}-\sqrt{2} \ket{2} \ket{4'}-\sqrt{2} \ket{3} \ket{3'}+\sqrt{2} \ket{4} \ket{2'} \\
-\sqrt{2} \ket{1} \ket{1'}-\sqrt{3} \ket{3} \ket{4'}-\sqrt{3} \ket{4} \ket{3'} \\
\sqrt{3} \ket{1} \ket{2'}+\sqrt{2} \ket{2} \ket{1'}+\sqrt{3} \ket{3} \ket{5'} \\
\sqrt{3} \ket{2} \ket{2'}+\sqrt{2} \ket{3} \ket{1'}+\sqrt{3} \ket{4} \ket{5'} \\
-\sqrt{3} \ket{1} \ket{4'}-\sqrt{3} \ket{2} \ket{3'}-\sqrt{2} \ket{4} \ket{1'}
\end{array}\right)_{\g{5}_1} \\
&\oplus \left(\begin{array}{c}
2 \ket{1} \ket{5'}+4 \ket{2} \ket{4'}+4 \ket{3} \ket{3'}+2 \ket{4} \ket{2'} \\
4 \ket{1} \ket{1'}+2 \sqrt{6} \ket{2} \ket{5'} \\
-\sqrt{6} \ket{1} \ket{2'}+2 \ket{2} \ket{1'}-\sqrt{6} \ket{3} \ket{5'}+2 \sqrt{6} \ket{4} \ket{4'} \\
2 \sqrt{6} \ket{1} \ket{3'}-\sqrt{6} \ket{2} \ket{2'}+2 \ket{3} \ket{1'}-\sqrt{6} \ket{4} \ket{5'} \\
2 \sqrt{6} \ket{3} \ket{2'}+4 \ket{4} \ket{1'}
\end{array}\right)_{\g{5}_2}
\\
\matc{
\ket{1}\\
\ket{2}\\
\ket{3}\\
\ket{4}\\
\ket{5}
}_{\g{5}}
\otimes
\matc{
\ket{1'}\\
\ket{2'}\\
\ket{3'}\\
\ket{4'}\\
\ket{5'}
}_{\g{5}} &= (\ket{1} \ket{1'}+\ket{2} \ket{5'}+\ket{3} \ket{4'}+\ket{4} \ket{3'}+\ket{5} \ket{2'})_{\g{1}_S} \\
&\oplus \left(\begin{array}{c}
\ket{2} \ket{5'}+2 \ket{3} \ket{4'}-2 \ket{4} \ket{3'}-\ket{5} \ket{2'} \\
-\sqrt{3} \ket{1} \ket{2'}+\sqrt{3} \ket{2} \ket{1'}+\sqrt{2} \ket{3} \ket{5'}-\sqrt{2} \ket{5} \ket{3'} \\
\sqrt{3} \ket{1} \ket{5'}+\sqrt{2} \ket{2} \ket{4'}-\sqrt{2} \ket{4} \ket{2'}-\sqrt{3} \ket{5} \ket{1'}
\end{array}\right)_{\g{3}_A} \\
&\oplus \left(\begin{array}{c}
2 \ket{2} \ket{5'}-\ket{3} \ket{4'}+\ket{4} \ket{3'}-2 \ket{5} \ket{2'} \\
\sqrt{3} \ket{1} \ket{3'}-\sqrt{3} \ket{3} \ket{1'}+\sqrt{2} \ket{4} \ket{5'}-\sqrt{2} \ket{5} \ket{4'} \\
-\sqrt{3} \ket{1} \ket{4'}+\sqrt{2} \ket{2} \ket{3'}-\sqrt{2} \ket{3} \ket{2'}+\sqrt{3} \ket{4} \ket{1'}
\end{array}\right)_{\g{3'}_A}\\
&\oplus \left(\begin{array}{l}
3 \sqrt{2} \ket{1} \ket{2'}+3 \sqrt{2} \ket{2} \ket{1'}-\sqrt{3} \ket{3} \ket{5'}+4 \sqrt{3} \ket{4} \ket{4'}-\sqrt{3} \ket{5} \ket{3'} \\
3 \sqrt{2} \ket{1} \ket{3'}+4 \sqrt{3} \ket{2} \ket{2'}+3 \sqrt{2} \ket{3} \ket{1'}-\sqrt{3} \ket{4} \ket{5'}-\sqrt{3} \ket{5} \ket{4'} \\
3 \sqrt{2} \ket{1} \ket{4'}-\sqrt{3} \ket{2} \ket{3'}-\sqrt{3} \ket{3} \ket{2'}+3 \sqrt{2} \ket{4} \ket{1'}+4 \sqrt{3} \ket{5} \ket{5'} \\
3 \sqrt{2} \ket{1} \ket{5'}-\sqrt{3} \ket{2} \ket{4'}+4 \sqrt{3} \ket{3} \ket{3'}-\sqrt{3} \ket{4} \ket{2'}+3 \sqrt{2} \ket{5} \ket{1'}
\end{array}\right)_{\g{4}_S} \\
&\oplus \left(\begin{array}{c}
\sqrt{2} \ket{1} \ket{2'}-\sqrt{2} \ket{2} \ket{1'}+\sqrt{3} \ket{3} \ket{5'}-\sqrt{3} \ket{5} \ket{3'} \\
-\sqrt{2} \ket{1} \ket{3'}+\sqrt{2} \ket{3} \ket{1'}+\sqrt{3} \ket{4} \ket{5'}-\sqrt{3} \ket{5} \ket{4'} \\
-\sqrt{2} \ket{1} \ket{4'}-\sqrt{3} \ket{2} \ket{3'}+\sqrt{3} \ket{3} \ket{2'}+\sqrt{2} \ket{4} \ket{1'} \\
\sqrt{2} \ket{1} \ket{5'}-\sqrt{3} \ket{2} \ket{4'}+\sqrt{3} \ket{4} \ket{2'}-\sqrt{2} \ket{5} \ket{1'}
\end{array}\right)_{\g{4}_A}\\
&\oplus \left(\begin{array}{c}
2 \ket{1} \ket{1'}+\ket{2} \ket{5'}-2 \ket{3} \ket{4'}-2 \ket{4} \ket{3'}+\ket{5} \ket{2'} \\
\ket{1} \ket{2'}+\ket{2} \ket{1'}+\sqrt{6} \ket{3} \ket{5'}+\sqrt{6} \ket{5} \ket{3'} \\
-2 \ket{1} \ket{3'}+\sqrt{6} \ket{2} \ket{2'}-2 \ket{3} \ket{1'} \\
-2 \ket{1} \ket{4'}-2 \ket{4} \ket{1'}+\sqrt{6} \ket{5} \ket{5'} \\
\ket{1} \ket{5'}+\sqrt{6} \ket{2} \ket{4'}+\sqrt{6} \ket{4} \ket{2'}+\ket{5} \ket{1'}
\end{array}\right)_{\g{5}_{S,1}} \\
&\oplus \left(\begin{array}{c}
2 \ket{1} \ket{1'}-2 \ket{2} \ket{5'}+\ket{3} \ket{4'}+\ket{4} \ket{3'}-2 \ket{5} \ket{2'} \\
-2 \ket{1} \ket{2'}-2 \ket{2} \ket{1'}+\sqrt{6} \ket{4} \ket{4'} \\
\ket{1} \ket{3'}+\ket{3} \ket{1'}+\sqrt{6} \ket{4} \ket{5'}+\sqrt{6} \ket{5} \ket{4'} \\
\ket{1} \ket{4'}+\sqrt{6} \ket{2} \ket{3'}+\sqrt{6} \ket{3} \ket{2'}+\ket{4} \ket{1'} \\
-2 \ket{1} \ket{5'}+\sqrt{6} \ket{3} \ket{3'}-2 \ket{5} \ket{1'}
\end{array}\right)_{\g{5}_{S,2}} 
\end{align*}
\endgroup
\section{Details of the fit} \label{app:fit}
The optimization problem of finding a good benchmark point for the model is based on minimizing the following cost function:
\begin{eqnarray} \label{appeq:X2}
   \chi^2= \chi^2_{O}+\chi^2_{c}= \sum_k  \left(\frac{\langle O_k\rangle-\hat{O_k}}{\sigma_{O_k}}\right)^2 + \sum_i \left( \frac{|c_i|-1}{\sigma_c}\right)^2,
\end{eqnarray}
where $O_k$ stands for the observable with measured value $\langle O_k\rangle \pm \sigma_{O_k}$ and model prediction $\hat{O}_k$.
In our case, $k \in [1,\, 4]$ runs over the three lepton masses and the anomalous magnetic moment of the muon.
The $\chi^2_c$ implies a normal distribution for the absolute value of the coefficients $c_i$ with standard deviation $\sigma_{c_i}$.
It is a prior that expresses the requirement that a successful flavor model should feature non hierarchical coefficients and alleviates over-fitting.

The fine tuning of the model is determined as \cite{Fedele:2020fvh}:
\begin{eqnarray} \label{eq:FTfunct}
    \Delta_{\rm FN}\equiv{\rm Max}_{k,i}|\delta_{K,i}|, \qquad \delta_{K,i}\equiv\frac{c_i}{\hat{O}_k}\cfrac{\delta \hat{O}_k}{\delta c_i}.
\end{eqnarray}
where $K\in [1,4]$ and $i\in [1,7]$ run over the number of observables and parameters, respectively.

\bibliography{references}
\newpage
\bibliographystyle{JHEP}

\end{document}